\documentclass[preprint,12pt,authoryear]{elsarticle}
\usepackage{amsmath, amsthm}
\usepackage{graphicx}
\usepackage{geometry}
\usepackage{enumitem}
\usepackage{float}
\usepackage{booktabs}
\usepackage{tabularx}
\usepackage{algorithm}
\usepackage{tikz}
\usetikzlibrary{positioning,shapes,arrows,arrows.meta,calc}
\usepackage{amsfonts}
\newcommand{\E}{\mathbb{E}}

\usepackage{algpseudocode}
\usepackage{amssymb}

\usepackage{bm}
\usepackage[bottom]{footmisc}

\usepackage{subcaption}
\usepackage{caption}
\usepackage{xcolor}
\usepackage[%
  colorlinks  = true,
  linkcolor   = {blue!50!black},
  urlcolor    = blue,
  citecolor   = {blue!50!black},
  linktocpage = true,
]{hyperref}
\usepackage[]{natbib}
\definecolor{darkblue}{RGB}{0,0,139}
\newcommand{\bt}[1]{\textcolor{darkblue}{#1}}
\newtheorem{proposition}{Proposition}

\title{Generative Learner for Distributional Causal Effects}

\author[mn]{Maria Nareklishvili\corref{cor1}}
\ead{marnar@stanford.edu}
\author[np]{Nicholas Polson}
\ead{ngp@chicagobooth.edu}
\author[vs]{Vadim Sokolov}
\ead{vsokolov@gmu.edu}

\cortext[cor1]{Corresponding author. Second Affiliation: University of Glasgow, Adam Smith Business School, Glasgow, United Kingdom. Replication code: \href{https://github.com/marnare/generativeAI/tree/main}{\texttt{marnare/generativeAI}}.  The authors thank NISS-CANSSI Collaborative Data Science workshop participants for valuable feedback and comments. }
\address[mn]{Graduate School of Business, Stanford University, Palo Alto, United States; marnar@stanford.edu}
\address[np]{Booth School of Business, University of Chicago, Chicago, United States; ngp@chicagobooth.edu}
\address[vs]{Department of Systems Engineering and Operations Research, George Mason University, Fairfax, United States; vsokolov@gmu.edu}

\begin{document}

\begin{abstract}
\noindent
We propose a generative learner for estimating conditional average treatment effects and characterizing the full distribution of these effects. The learner takes the form of a multi-head feed-forward neural network with three jointly estimated subnetworks: the propensity score, the conditional quantile function of the baseline outcome, and the conditional quantile treatment effect. Each quantile-indexed subnetwork combines a covariate representation with cosine quantile embeddings through element-wise multiplication. We then recover the conditional average treatment effect as an integral over conditional quantile treatment effects. Under the classical causal assumptions within the Neyman--Rubin potential outcomes framework, we find that the proposed generative learner reduces out-of-sample mean squared error relative to the generalized random forest, double machine learning, and generative adversarial networks, with gains ranging from 5.4\% to 93.5\% on average across experimental designs. In an empirical application, we formalize the Stefan--Boltzmann law within a unidirectional causal model and apply the method to publicly available stellar data. The estimated effects satisfy the restrictions the law implies.
\end{abstract}


\maketitle

\section{Introduction}

Estimating conditional average treatment effects (CATEs) from observational
data is a central problem in causal inference. In the potential outcomes
framework \citep{rubin1974estimating, rubin2005causal}, the causal effect for a
unit is the difference between its outcomes under treatment and under control.
Only one of the two is ever observed, which is the fundamental missing data
problem. A large literature now tackles this problem with flexible machine
learning methods. Double/debiased machine learning
\citep[DML;][]{chernozhukov2018double} estimates nuisance functions with
generic learners and uses cross-fitting and orthogonal scores to retain valid
inference on the treatment effect. Bayesian causal forests
\citep[BCF;][]{hahn2020bayesian} split the conditional mean into a prognostic
component and a treatment-effect component, each modeled by its own Bayesian
Additive Regression Tree (BART) ensemble. Both target the CATE, the expected
difference in potential outcomes among units with the same observed
characteristics.

Treatment effects, however, can vary in two distinct ways: across units with
different covariates, and across the outcome distribution itself. Capturing
both in one model is hard, and existing methods tend to handle one at the
expense of the other. The generalized random forest partitions the covariate
space and estimates an average contrast within each neighborhood
\citep{athey2019generalized}, and the DML literature targets conditional or
unconditional mean effects \citep{chernozhukov2018double}. These methods let
the effect vary with covariates but say nothing about where in the outcome
distribution it falls. Quantile treatment effect estimators do the reverse:
they describe the effect at different points of the outcome distribution, but
are usually evaluated at a few covariate values or require a separate fit for
each covariate of interest. The distinction matters whenever the question is
not only who is affected but also which part of the outcome distribution the
effect reaches, as with clinical interventions on patients whose baseline
health varies, or policies whose distributional consequences are the main
concern.

Our approach builds on the quantile representation of the conditional mean
discussed by \citet{parzen2004quantile}. Provided the relevant first moments
exist, a conditional mean equals the integral of the corresponding conditional
quantile function over the unit interval. The CATE is therefore the difference
between the integrated conditional quantile functions of the treated and
control potential outcomes, and it can be recovered from a richer object: the
pair of conditional quantile functions themselves. To estimate these, we
introduce a \textit{counterfactual outcome generator} within the
Neyman--Rubin potential outcomes framework. The generator maps a covariate profile, a treatment
state, and an independent uniform noise variable to the corresponding
potential outcome. Its existence follows from the noise-outsourcing (transfer)
theorem of \citet{kallenberg1997foundations}, which states that a conditional
distribution can be written as a measurable function of the conditioning
variables and an independent source of randomness. Such representations are
not unique in general, but there is a canonical choice: take the noise to be
uniform on $(0,1)$ and require the generator to be non-decreasing and
left-continuous in it. The generator is then the conditional quantile function
of the potential outcome, so estimating the generator amounts to estimating
the conditional quantile functions of the potential outcomes.

Our main contribution is GenTE,\footnote{GenTE stands for generative treatment
effect learner.} a multi-head feed-forward neural network that estimates three
functions jointly. Two of them come from the potential outcomes framework
\citep{rubin1974estimating, rubin2005causal}: the baseline, which describes
the outcome under control, and the treatment effect, the difference in
expected potential outcomes for units with the same covariates. The third is
the propensity score, the probability of treatment given covariates. It
describes how treatment is assigned rather than how outcomes are generated, and
we include it because assignment in observational data may depend on
covariates. What sets the architecture apart is that both the baseline and the
treatment effect depend on a quantile level as well as on the covariates. The
baseline is the conditional quantile function of the control outcome, and the
treatment-effect subnetwork returns the conditional quantile treatment effect
at the same level. Following \citet{dabney2018implicit}, both take a
\textit{compositional form}: learned covariate features are multiplied
element-wise by a cosine embedding of the quantile level, and the two
subnetworks share that embedding. The treatment effect can thus vary across
covariate profiles and across the outcome distribution at the same time,
without discretizing either. Because a conditional mean is the integral of
the conditional quantile function \citep{parzen2004quantile}, the CATE follows
by integrating the estimated effect over the quantile level.

All three components are estimated by minimizing a single objective with two
main terms. The first is a pinball loss on the observed outcome. Its expected
value is minimized at the conditional quantile, so it is what identifies the
quantile functions targeted by the baseline and treatment-effect subnetworks.
The second is a binary cross-entropy loss on the treatment indicator, which
estimates the assignment mechanism. Since the propensity subnetwork passes an
embedding to the baseline, this term also pushes the baseline's input to carry
information about who is treated. A further penalty discourages quantile
crossing, so the fitted functions are increasing in the quantile level, as any
valid quantile function must be. The quantile level is redrawn throughout
training, which means the network learns the whole unit interval rather than a
fixed set of quantiles, and one fitted model gives the treatment effect at any
combination of covariates and quantile level. Joint estimation also differs
from two-stage procedures, where the propensity score is fitted first and then
held fixed. In GenTE, the propensity subnetwork receives gradients from both
the assignment and the outcome terms, so the representation it passes to the
baseline reflects the outcome model as well and keeps changing during training.

Three design choices shape the architecture. First, each latent quantity (the
quantile-specific baseline, the quantile-specific treatment effect, and the
propensity score) has its own subnetwork. The treatment effect is therefore an
explicit part of the model, not something recovered afterwards as the
difference between two separately estimated response surfaces, as in a
T-learner. With a T-learner, any bias in either surface carries straight into
the difference, and nothing in the estimation makes the two biases cancel.
Second, the baseline is a composition of covariate features and the quantile
embedding \citep{dabney2018implicit}, so it is a conditional quantile function
rather than a conditional mean and can capture heterogeneity across the
distribution of the control outcome. The propensity subnetwork also supplies
the baseline encoder with a learned representation of the assignment
mechanism, letting the baseline absorb systematic variation in treatment
propensity, in the spirit of \citet{hahn2020bayesian}. Third, the
treatment-effect subnetwork has the same compositional structure instead of
being an unstructured function of $(x, q)$: covariate features are multiplied
element-wise by a quantile embedding built from the cosine frequencies
$\cos(\pi i q)$, $i = 0, \dots, n-1$, passed through a learned linear layer and
a rectified linear unit (ReLU). When the target functions are themselves
compositional, this is the setting in which deep ReLU estimators achieve
convergence rates governed by the intrinsic rather than the ambient covariate
dimension (Propositions~\ref{prop:sh} and~\ref{prop:padilla}). Estimation is
an ordinary supervised learning problem: given outcomes, treatments, and
covariates, the network is fitted by gradient-based optimization, with no
Markov chain Monte Carlo (MCMC) or explicit posterior computation.

We compare GenTE with three widely used methods in causal machine learning:
the generalized random forest (GRF) \citep{athey2019generalized}, a
residual-on-residual estimator within the double/debiased machine learning
framework \citep{chernozhukov2018double} (DML--XGB), and GANITE, a generative
adversarial network for individualized treatment effects
\citep{yoon2018ganite}. Across a broad set of Monte Carlo designs, GenTE
attains the lowest out-of-sample mean squared error in 30 of 36 cases, and
its own accuracy is stable across noise distributions and treatment assignment
mechanisms. As an empirical application, we cast the Stefan--Boltzmann law as
a unidirectional causal model in which temperature determines stellar
luminosity. On publicly available stellar data, the estimates satisfy the
restrictions the law implies: the effect of surface temperature on luminosity
is positive throughout, largest for supergiants and hypergiants, and smallest
for dwarfs.

The paper is organized as follows. Section~2 reviews related work. Section~3
sets out the causal model and the identification assumptions. Section~4
describes the architecture of the generative learner. Section~5 presents the
Monte Carlo simulations, and Section~6 turns to the Stefan--Boltzmann law.
Section~7 discusses the results and directions for future work.

\section{Related Literature}
Causal machine learning is an emerging field that has expanded rapidly in recent years. \citet{athey2016recursive} adapt conventional classification and regression trees (CART) to the estimation of conditional average treatment effects by introducing a mean squared error criterion tailored to treatment effect estimation. \citet{wager2018estimation, lechner2018modified,  athey2019estimating, athey2019generalized, lechner2022modified} extend the random forest framework to estimate a range of causal functionals, including the average treatment effect and local average treatment effect, and establish asymptotic properties of the resulting estimators. A central modification in this literature is the use of “honest” trees and forests: the trees and forest are trained on one subsample of the data, while treatment effects are estimated on a separate, held-out subsample, thereby reducing bias from in-sample prediction.  

The second related class of estimators, and the current gold standard for conditional average treatment effect estimation, belongs to the double/debiased machine learning (DML) literature. \citet{chernozhukov2016double, chernozhukov2018double} generalize the estimators developed in the semiparametric statistics literature to accommodate a broad class of machine learning models for estimating nuisance parameters (propensity score and the outcome model). Estimators in this class fit nuisance models on one subsample and use them to construct treatment effect estimates on a separate sample. The authors propose estimators that are consistent, asymptotically efficient, and inherit desirable large sample properties from semiparametric statistics (see \cite{ahrens2025introduction} for further review on doubly-robust estimators). 

The third related category of estimators of treatment effects are based on deep learners. \cite{farrell2018deep} investigate theoretical properties of deep neural networks for estimating conditional average treatment effects. \cite{berrevoets2023causal} use deep learners for investigating structural relationships in the data.  Generative learner in this paper is also broadly related to modeling parameters with flexible machine learning models \citep{nareklishvili2023deep, polson2018posterior, polson2025generative, polson2024generative, polson2024generativea}.

Bayesian tree-based methods constitute another important class of conditional average treatment effect estimators. \citet{hill2011bayesian} applies Bayesian Additive Regression Trees (BART) to causal inference, leveraging flexible tree ensembles with regularization priors to estimate individual-level treatment effects. \citet{hahn2020bayesian} introduce Bayesian causal forests (BCF), which explicitly decompose the conditional mean into a prognostic component and a treatment-effect component modeled by separate tree ensembles, a structural decomposition that shares conceptual motivation with the multi-head architecture proposed here. \citet{papakostas2024deep} implement the BCF decomposition using fully separate neural networks and find that this architecture outperforms shared-hidden-layer designs, especially when treatment effects are small relative to prognostic effects. \citet{mcculloch2021causal} extend BART to instrumental variable settings with Dirichlet process mixture error distributions, and \citet{alcantara2025learning} adapt BART for regression discontinuity designs with linear leaf models.

The compositional cosine embedding employed in our treatment-effect subnetwork connects to a body of work on neural network approximation theory.  \citet{barron1993universal} shows that single-hidden-layer networks achieve an integrated squared approximation error of order $O(C^2/N)$, where $C$ is the first absolute moment of the Fourier transform of the target function and $N$ is the number of hidden units, a rate that is independent of input dimension and thus avoids the curse of dimensionality. \citet{coppejans2000breaking} builds on Kolmogorov's superposition theorem \citep{coppejans2004kolmogorovs} to propose nonparametric estimators that express multivariate regression functions as compositions of univariate functions, providing an early econometric argument for why compositional architectures, such as deep neural networks, can circumvent the curse of dimensionality. \citet{schmidt-hieber2020nonparametric} extends these ideas to deep ReLU networks and establishes minimax-optimal rates for nonparametric regression. On the distributional side, \citet{parzen2004quantile} argues that the quantile function provides a more complete and efficient basis for statistical modeling than the mean, since $\mathbb{E}[Y \mid X] = \int_0^1 F_{Y|X}^{-1}(q)\,dq$,  that is, the mean is a special case of the quantile integral. Quantile neural networks, first proposed by \citet{white1992nonparametric} and extended to high-dimensional settings by \citet{padilla2022quantile} and \citet{shen2021deep}, provide the foundation for learning conditional quantile functions with flexible architectures. The cosine embedding we adopt is closely related to the implicit quantile network (IQN) of \citet{dabney2018implicit}, which parameterizes the quantile function via a learned linear transformation of cosine basis features $\cos(\pi i q)$ for $i = 0,\dots,n-1$, enabling a fully nonparametric representation of conditional distributions. The implicit quantile network of \citet{dabney2018implicit} generalizes
the quantile regression approach of \citet{dabney2018distributional},
which learns a fixed set of quantile locations, to a continuous representation over the entire quantile domain.

Our approach is also closely related to GANITE \citep{yoon2018ganite}, which applies generative adversarial networks to estimate CATEs. GANITE uses a generative adversarial network (GAN) to impute counterfactual outcomes, i.e., to generate each unit's unobserved potential outcome under the treatment state not realized in the data, while a discriminator is trained to make the imputed outcomes statistically indistinguishable from observed outcomes. In a subsequent stage, a separate predictor network maps covariates to individualized treatment effects using the completed set of potential outcomes (observed plus imputed). GenTE is a single-stage estimator: the outcome components and the propensity score are learned jointly by minimizing a composite loss, without a separate counterfactual-imputation step. By contrast, GANITE follows a two-stage pipeline, imputation via a GAN followed by supervised prediction of treatment effects. The imputed counterfactuals enter
the second stage as if they were data, so first-stage error is carried
forward as measurement error in the regressand and the loss function
carries no information on how the imputation is trained.  In GenTE, the
effect enters as its own component, so it is regularized directly rather
than left as the difference of two separately fitted surfaces, and the
propensity score is learned in the same representation, so the
assignment mechanism continues to shape the outcome model rather than
being fixed in an earlier stage that later training cannot revise.

\section{Causal Model}

\subsection{Counterfactual Outcomes }\label{subs_identification}

We work within the Neyman--Rubin potential outcomes framework
\citep{rubin1974estimating, rubin2005causal}. Consider $S$ observational units indexed by
$s = 1,\dots,S$, drawn independently and identically from a distribution
\[
\bigl(Y_s(1),\,Y_s(0),\,D_s,\,X_s\bigr) \sim \mathbb{P}_0 ,
\]
where $X_s \in \mathbb{R}^p$ is a vector of covariates, and $D_s \in \{0,1\}$ is a binary
treatment indicator. For each treatment state \(d \in \{0, 1\}\), let \(Y_s(d)\) denote the outcome unit \(s\) would realize if assigned \(D_s = d\). Thus, $Y_s(1)$ and  $Y_s(0)$ are the treated and control potential (counterfactual) outcomes, respectively. Only one potential outcome is realized per unit in practice, so the
observed outcome is 
\begin{equation}
Y_s = D_s Y_s(1) + (1-D_s) Y_s(0).
\label{eq:switching}
\end{equation}

It is convenient to separate each potential outcome into a component determined by
covariates and an idiosyncratic component. Define the baseline conditional mean and
the conditional average treatment effect (CATE),
\[
\mu(x) := \mathbb{E}\bigl[Y(0) \mid X = x\bigr],
\qquad
\theta(x) := \mathbb{E}\bigl[Y(1) - Y(0) \mid X = x\bigr],
\]
so that
\begin{align}
Y_s(0) &= \mu(X_s) + \varepsilon_s(0), \label{eq:y0}\\
Y_s(1) &= \mu(X_s) + \theta(X_s) + \varepsilon_s(1), \label{eq:y1}
\end{align}
where $\varepsilon_s(d) := Y_s(d) - \mathbb{E}[Y_s(d) \mid X_s]$ satisfies
$\mathbb{E}[\varepsilon_s(d) \mid X_s] = 0$ for $d \in \{0,1\}$ by construction.
Apart from the existence of the relevant conditional first moments,  \eqref{eq:y0}--\eqref{eq:y1} are definitional and impose no functional-form restriction on
$\mathbb{P}_0$: the functions $\mu(\cdot)$ and $\theta(\cdot)$ are left
unspecified, and the disturbances may be heteroskedastic.  The same decomposition is described by equation (8) of \cite{heckman1998characterizing} where the authors assume separability between a flexible function of covariates and the  error terms.


The unit-level treatment effect
\[
\theta_s := Y_s(1) - Y_s(0) = \theta(X_s) + \varepsilon_s(1) - \varepsilon_s(0)
\]
is never observed, since each unit is observed in only one treatment state.
Substituting \eqref{eq:y0}--\eqref{eq:y1} into \eqref{eq:switching} yields
\begin{equation}
Y_s = \mu(X_s) + D_s\,\theta(X_s) + \varepsilon_s,
\qquad
\varepsilon_s := D_s\varepsilon_s(1) + (1-D_s)\varepsilon_s(0).
\label{eq:observed}
\end{equation}

A parametric version of \eqref{eq:observed} is derived by equation (2) of \cite{heckman1990varieties}. To nonparametrically identify the conditional distributions of the potential
outcomes, and hence the CATE from observational data, we impose the following
standard sufficient conditions \citep{rubin2005causal}:
\begin{itemize}
\item[$\circ$] Stable unit treatment value assumption (SUTVA): each unit's observed outcome depends
only on its own treatment assignment, ruling out interference between units and multiple versions of the treatment. 
\item[$\circ$] Overlap: $0 < \pi(x) < 1$ for all $x$ in the support of $X$,
where $\pi(x) := \Pr(D = 1 \mid X = x)$: units with covariates $x$ may appear in
either treatment state.
\item[$\circ$] Conditional unconfoundedness $D_s \perp\!\!\!\perp \bigl(Y_s(1), Y_s(0)\bigr) \mid X_s$:
conditional on covariates, treatment assignment is independent of potential outcomes.
\end{itemize}

Together these assumptions imply $F_{Y(d)\mid X = x}(y) = F_{Y \mid X = x,\, D = d}(y)$ for almost all \(x\) in the support of \(X\) and  \(y\) in the relevant outcome support with
$d \in \{0,1\}$, so that the conditional distribution of each potential outcome, and
hence $\theta(x)$, is recoverable from the observed data. This equality is the basis
for the quantile-based construction developed in Section \ref{sec_gente}.

\subsection{Treatment assignment mechanism}\label{sec_treat_assign}
Because the units are observational rather than experimental, treatment is not
necessarily randomly assigned: $D_s$ may depend on $X_s$. We make the assignment mechanism
explicit through a classical threshold-crossing representation,
\begin{equation}
D_s = \mathbf{1}\bigl\{\, \nu(X_s) > u_s \,\bigr\},
\label{eq:selection}
\end{equation}
where $\nu:\mathbb{R}^p \to \mathbb{R}$ is an unknown selection index and $u_s$ is a
latent disturbance, independent of $X_s$, with continuous and strictly increasing
distribution function $F_u$ on its support. \(\mathbf{1}\{\cdot\}\) is a binary indicator for \(\nu(X_s) > u_s\). The threshold-crossing representation in \eqref{eq:selection} is the latent-index
selection model of \citet{heckman1974shadow, heckman1979sample, heckman2005structural}.
The propensity score is then
\[
\pi(x) \;=\; \Pr\bigl(D = 1 \mid X = x\bigr) \;=\; F_u\bigl(\nu(x)\bigr).
\]
The threshold-crossing representation by itself imposes no restriction on the conditional treatment probability.  Two points are worth emphasizing. First, $\nu(\cdot)$ is unrestricted, so selection
into treatment may depend on covariates in an arbitrary, possibly nonlinear way. Unlike simple randomized experiments with a constant assignment probability, observational treatment assignment may depend on covariates in an arbitrary, potentially nonlinear manner. Second, the identifying content
lies not in \eqref{eq:selection} itself but in how $u_s$ relates to the outcome
disturbances. A sufficient condition for conditional unconfoundedness to hold is
\begin{equation}
u_s \;\perp\!\!\!\perp\; \bigl(\varepsilon_s(0),\, \varepsilon_s(1)\bigr) \;\big|\; X_s.
\label{eq:selection-on-observables}
\end{equation}
That is, selection should be driven by covariates together with a shock unrelated to
the unobserved determinants of outcomes, i.e., selection on observable covariates. Condition
\eqref{eq:selection-on-observables} is sufficient for unconfoundedness, since
conditional on $X_s$ the treatment indicator is a function of $u_s$ alone and the
potential outcomes are functions of $\bigl(\varepsilon_s(0), \varepsilon_s(1)\bigr)$
alone. If instead \(u_s\) and the potential-outcome disturbances are dependent
conditional on \(X_s\), for example because units select on anticipated gains
that are not captured by \(X_s\), then unconfoundedness may fail, and $\theta(\cdot)$ is not generally identified from observational data under the maintained assumptions. In that case, identification would require additional structure, such as a
valid instrumental variable, a control-function approach, or another
quasi-experimental design \citep{ angrist1996identification,
blundell2003endogeneity}.

\subsection{Relation to DAGs}

An alternative representation of the identifying structure is provided by the
causal graphical framework of \citet{pearl2009causality}. Writing $X$, $D$, and
$Y$ for a generic draw from $\mathbb{P}_0$, the causal relations among the
observed variables in the setting considered here can be represented
schematically as
\[
X \rightarrow D, \qquad X \rightarrow Y, \qquad D \rightarrow Y,
\]
where $X$ denotes a collection of pre-treatment covariates. This representation
suppresses possible causal relations among the individual components of $X$,
which are not required for the adjustment argument below.

In the graphical formulation, identification follows if $X$ blocks all
back-door paths from treatment to outcome. That is, after conditioning on
$X$, there must be no remaining open path that induces confounding between $D$
and $Y$. Provided that this condition holds and that
$0<\pi(x)<1$, Pearl's back-door criterion implies
\[
F_{Y\mid \operatorname{do}(D=d),\,X=x}(y)
=
F_{Y\mid D=d,\,X=x}(y).
\]
Under the usual correspondence between interventions and potential outcomes,
the interventional distribution satisfies
\[
F_{Y\mid \operatorname{do}(D=d),\,X=x}(y)
=
F_{Y(d)\mid X=x}(y),
\]
so that
\[
F_{Y(d)\mid X=x}(y)
=
F_{Y\mid D=d,\,X=x}(y).
\]
This is the identification relation obtained above from overlap and
conditional unconfoundedness.  Importantly, the DAG encodes causal relations
rather than functional form restrictions. Interactions among the components of
$X$ may enter the treatment assignment and outcome mechanisms through the
unrestricted functions $\nu(x)$, $\mu(x)$, and $\theta(x)$, and are therefore
accommodated by the flexible neural-network parameterization. A richer
graphical formulation could additionally model or learn the causal relations
among the components of $X$ and account for uncertainty about the underlying
graph, as in Bayesian approaches to causal structure learning and causal effect
estimation \citep{castelletti2020discovering,castelletti2024joint}.

\section{Generative Causal Learner}\label{sec_gente}

We approach estimation of CATEs through a counterfactual outcome generator rather than through the
conditional means directly. Let $U \sim \mathrm{Unif}(0,1)$ be a uniformly distributed reference noise
variable, independent of $(X_s, D_s)$, and consider a measurable function $G$ (counterfactual outcome generator) that returns a
potential outcome from a covariate value, a treatment state, and a draw of this noise,
\begin{equation}
\bigl(Y_s(d) \mid X_s = x\bigr) \;\stackrel{d}{=}\; G\bigl(U,\, x,\, d\bigr),
\qquad d \in \{0,1\},
\label{eq:cf-generator}
\end{equation}
where \(\stackrel{d}{=}\) denotes equality in distribution. Intuitively, a counterfactual outcome generator is a deterministic map that, evaluated at uniform noise, reproduces the conditional distribution of each potential outcome given covariates. The noise-outsourcing theorem of \citet[Theorem 5.10]{kallenberg1997foundations}, commonly referred to as the transfer theorem, 
guarantees that such
a map exists: a conditional distribution can be expressed as a deterministic function
of its conditioning inputs and an independent source of randomness, so
\eqref{eq:cf-generator} assumes nothing beyond what the joint law $\mathbb{P}_0$
already provides.  The canonical non-decreasing, left-continuous choice is the conditional quantile function:
\begin{equation}
G(q,\, x,\, d) \;=\; F^{-1}_{Y(d) \mid X = x}(q), \qquad q \in (0,1).
\label{eq:generator-is-quantile}
\end{equation}
Overlap and conditional unconfoundedness together with SUTVA deliver the identification equality
\[F_{Y(d) \mid X = x}(y) = F_{Y \mid X = x,\, D = d}(y),\] so the generator is estimable from
observed triples $\{(Y_s, X_s, D_s)\}_{s=1}^{S}$ as
\[G(q,x,d) = F^{-1}_{Y \mid X = x,\, D = d}(q).\] 

\noindent The causal identification assumptions therefore allow us to move from the
unobserved potential outcome distributions to the corresponding observed
conditional outcome distributions. Estimating the counterfactual generator is
then equivalent to estimating the treatment-specific conditional quantile
functions, from which the conditional quantile treatment effects and,
by integration, the CATE are obtained. 

An alternative would be to specify a generative model for the joint
counterfactual outcome distribution and use it to simulate a sufficiently large sample
of paired potential outcomes,
\[
\bigl(Y_m(0),Y_m(1),X_m\bigr), \qquad m=1,\ldots,M,
\]
where \(M\) is the total number of simulated draws. Each simulated pair would imply an individual treatment effect,
\[
\theta_m = Y_m(1)-Y_m(0).
\]
To construct an observable analogue of the simulated data, one could further
draw a treatment assignment $D_m$ and by \eqref{eq:switching} obtain synthetic tuples
\(
\bigl(X_m,D_m,Y_m,\theta_m\bigr).
\)
These simulated data could then be used to learn the conditional distribution
of the treatment effect given the observed outcome, treatment status, and
covariates, for example \(
p(\theta_m\mid Y_m,D_m,X_m).
\) Such an approach, however, requires a model for the joint distribution
of $(Y_m(0),Y_m(1))\mid X_m$, as well as a sufficiently large
number of simulated draws. This approach is more extensively discussed by \cite{polson2024generativea, polson2025generative}.

In this paper, the primary estimand of interest is the conditional average treatment
effect. For covariate value $x$, it is the difference of the
integrated conditional quantile functions of the treated and control potential outcomes,
\begin{align}
\theta(x) &= \mathbb{E}\bigl[Y(1) - Y(0) \mid X = x\bigr] \nonumber \\
&= \int_0^1 F^{-1}_{Y(1)|X=x}(q)\,dq \;-\; \int_0^1 F^{-1}_{Y(0)|X=x}(q)\,dq .
\label{eq:cate-quantile}
\end{align}
Under overlap and conditional unconfoundedness the
identification equality $F_{Y(d)|X=x}(y) = F_{Y|X=x,\,D=d}(y)$ applies, and the estimand
becomes a functional of the distribution of the observed data,
\begin{equation}
\theta(x) = \int_0^1 F^{-1}_{Y|X=x,\,D=1}(q)\,dq
\;-\; \int_0^1 F^{-1}_{Y|X=x,\,D=0}(q)\,dq
= \int_0^1 \bigl[G(q,x,1) - G(q,x,0)\bigr] dq .
\label{eq:cate-identified}
\end{equation}

\noindent This representation exploits the fact that the conditional mean equals an integral over conditional quantiles \citep{parzen2004quantile} and requires
$\mathbb{E}[|Y(d)| \mid X = x] < \infty$. Under conditional rank invariance,\footnote{Conditional rank invariance requires the same quantile level $q$ to index both $Y_s(0)$ and $Y_s(1)$ conditional on $X_s$.}
the common quantile index can additionally be interpreted as an individual’s latent rank across treatment states, thereby identifying the distribution of unit-level treatment effects through differences in the corresponding potential-outcome quantiles.

A key insight is that conditional quantile functions are inherently \emph{compositional}: the conditional quantile $Q(q; Y\mid X)$ can be written as $Q_Y\!\bigl(Q(q; F_Y(Y)\mid X)\bigr)$, a composition of the marginal quantile transform and a nonlinear regression on covariates \citep{parzen2004quantile, polson2025generative}.  This compositional structure, rather than a traditional additive expansion, motivates the deep network architecture that we adopt.  Following \citet{dabney2018implicit}, we represent each conditional quantile function as covariate-quantile multiplicative interaction: 
\begin{equation}\label{eq:composite}
F_{Y(d)|X=x}^{-1}(q) \;\approx\; g_{d}\!\bigl(\psi(x)\odot\varphi(q)\bigr),
\end{equation}
\noindent where $\psi\colon\mathbb{R}^{p}\to\mathbb{R}^{k}$ is a covariate network that maps the high-dimensional covariates~$x$ into a learned summary \citep{jiang2017learning}, $\varphi\colon[0,1]\to\mathbb{R}^{k}$ is a quantile embedding network (with \(k = 64\) here), $\odot$ denotes element-wise (Hadamard) multiplication, and $g_{d}$ is the output network for treatment state~$d$.  The quantile embedding uses a cosine basis
\begin{equation}\label{eq:cosine_embed}
\varphi_{j}(q) \;=\; \operatorname{ReLU}\!\Biggl(\sum_{i=0}^{n-1}\cos(\pi i\, q)\,w_{ij} + b_{j}\Biggr), \qquad j=1,\dots,k,
\end{equation}
\noindent with $n=32$ by default. \footnote{$n=32$ is one of the values
examined by \citet{dabney2018implicit} (Figure 5), who retain $n=64$ in
their preferred specification. As with other architectural
hyperparameters, $n$ should be selected by cross-validation on the
data set at hand.}  The Hadamard product $\psi(x)\odot\varphi(q)$ is the central architectural choice: it \emph{composes} the covariate representation with the quantile representation rather than concatenating or adding them, mirroring the natural compositional structure of the quantile function.  The network parameters (all weights and biases) are randomly initialized from a prior distribution (uniform), and for each forward pass we sample quantile levels $q$ uniformly from $[0,1]$, enabling approximation of the full conditional distribution. Additionally, the loss function in \ref{app:loss} incorporates a penalty term that discourages quantile crossing, promoting the monotonicity in $q$ required for valid conditional quantile functions.

The CATE estimand is then written as
\begin{align}\label{eq_cate_composition}
\theta(x)\;\approx  \;\mathbb{E}_{q}\!\left[g_{1}\!\bigl(\psi(x)\odot\varphi(q)\bigr)
   - g_{0}\!\bigl(\psi(x)\odot\varphi(q)\bigr)\right].
\end{align}
\noindent
The covariate network $\psi(x)$ acts as a learned summary statistic that compresses the input $X_s\in\mathbb{R}^{p}$ while retaining the information relevant to each modeling target \citep{jiang2017learning}.  This dimensionality reduction is a key advantage over tree-based and kernel-based methods, which must partition the full covariate space and therefore suffer from the curse of dimensionality when $p$ is large \citep{coppejans2000breaking}.   Another key advantage of the parameterization in ~\eqref{eq:composite} is that it represents the target functions as compositions of simpler maps---exactly the structure for which deep ReLU networks achieve minimax-optimal convergence rates that depend on the intrinsic dimensions of each compositional layer of the oracle function rather than the ambient dimension~$p$ (Propositions~\ref{prop:sh} and~\ref{prop:padilla}).  Estimators that impose an additive structure escape the curse of
dimensionality only by assuming it away, and are misspecified whenever
the target involves interactions. Estimators that impose no structure
must average locally over the full covariate space and attain rates that
deteriorate rapidly in~$p$.  Because quantile functions update as compositions by their very nature \citep{polson2025generative}, deep neural networks are natural and plausibly more efficient methods for quantile-based treatment effect estimation than kernel or tree-based methods \citep{dabney2018implicit}.

\subsection{Approximation-theoretic motivation}

The compositional structure of the GenTE architecture is motivated by two results on the approximation and estimation rates of deep ReLU networks for H\"older-smooth functions. Propositions \autoref{prop:sh} and \autoref{prop:padilla}  bound how fast a sparse deep ReLU network estimator converges to the
true function as the sample size $S$ grows. They show that when the target
is a composition of simpler, low-dimensional functions, the convergence rate
is governed by the smoothness and intrinsic dimension of those pieces rather
than by the number of covariates $p$. Consequently, provided the target function has a compositional
representation with low intrinsic dimension, the statistical cost of
additional covariates is governed by that intrinsic dimension rather than
by $p$, so deep ReLU estimators avoid the curse of dimensionality to
which tree- and kernel-based nonparametric estimators are subject.

\begin{proposition}[Deep ReLU regression; \citet{schmidt-hieber2020nonparametric}, Theorem~1]
\label{prop:sh}
Let the oracle function \(f_0\) belong to the compositional H\"older class
$\mathcal{H}(m,\mathbf{w},t,\beta,K)$, meaning
\[
f_{0}
=
\zeta_{m}\circ \zeta_{m-1}\circ\cdots\circ \zeta_{0},
\]
where $m\in\mathbb{N}$ is the number of compositions,
$\mathbf{w}=(w_{0},\ldots,w_{m+1})$ are the layer widths
($w_{0}=p$ is the covariate dimension and $w_{m+1}=1$),
$t=(t_{0},\ldots,t_{m})$ are the intrinsic dimensions
($t_{i}\le w_{i}$: each coordinate $\zeta_{i,j}$ of $\zeta_{i}$ depends on at
most $t_{i}$ arguments and lies in the H\"older class),
$\beta=(\beta_{0},\ldots,\beta_{m})$ are the H\"older exponents, and
$K>0$ bounds the H\"older norms and the domains.
Write
\[
\beta_{i}^{*}
=
\beta_{i}\prod_{\ell=i+1}^{m}\min\{\beta_{\ell}, 1\}
\]
for the effective smoothness at layer $i$.
Let $\hat{f}_{S}$ be a least-squares estimator over a class of sparse
deep ReLU networks with depth $L\asymp\log S$.
Then
\[
\mathbb{E}\bigl\|\hat{f}_{S}-f_{0}\bigr\|^{2}
\;\lesssim\;
\max_{i=0,\dots,m}\;
S^{-2\beta_{i}^{*}/(2\beta_{i}^{*}+t_{i})}
\;\log^{3}S,
\]
where  \(\mathbb{E}\bigl\|\hat{f}_{S}-f_{0}\bigr\|^{2}\) is the expectation of the squared L2-norm. The rate depends on the intrinsic dimensions of the compositional function $t_{i}$, not on the
ambient input dimension $p$.
In particular, for additive models $f_{0}(x)=\sum_{j=1}^{p}f_{j}(x_{j})$ it reduces to
$S^{-2\beta/(2\beta+1)}\log^{3}S$, which is minimax optimal up to
logarithmic factors and does not grow with the covariate dimension $p$.
\end{proposition}


\begin{proposition}[Quantile ReLU regression; \citet{padilla2022quantile}, Theorem~6]
\label{prop:padilla}
Let the conditional $q$-quantile $f_q^*$ belong to the compositional
H\"older class $\mathcal{H}(m,\mathbf{w},t,\beta,K)$, meaning

\[
f_q^*
=
\zeta_{m}\circ \zeta_{m-1}\circ\cdots\circ \zeta_{0},
\]
where $m\in\mathbb{N}$ is the number of compositions,
$\mathbf{w}=(w_{0},\ldots,w_{m+1})$ are the layer widths
($w_{0}=p$ is the covariate dimension and $w_{m+1}=1$),
$t=(t_{0},\ldots,t_{m})$ are the intrinsic dimensions
($t_{i}\le w_{i}$: each coordinate of $\zeta_{i}$ depends on at most $t_{i}$
arguments),
$\beta=(\beta_{0},\ldots,\beta_{m})$ are the H\"older exponents of those
coordinates, and $K>0$ bounds the H\"older norms and the domains.
Write
\[
\beta_{i}^{*}
=
\beta_{i}\prod_{\ell=i+1}^{m}\min(\beta_{\ell},  1)
\]
for the effective smoothness at layer $i$.
Suppose $\hat{f}$ minimizes the pinball loss over a class of sparse deep
ReLU networks with depth $L\asymp\log S$.
Assume that the conditional density of $Y\mid X$ is bounded above and
below at $f_q^*(X)$ (the error distribution may be heavy-tailed).
Then, with probability approaching one,
\[
\bigl\|\hat{f}-f_q^*\bigr\|_{S}^{2}
\;\le\;
C\;
\max_{i=0,\dots,m}\;
S^{-2\beta_{i}^{*}/(2\beta_{i}^{*}+t_{i})}
\;\log^{3}S,
\]
where $C>0$ is a constant and \(
\|h\|_{S}^{2}
=
\frac{1}{S}\sum_{s=1}^{S}h(X_{s})^{2}
\) is an empirical norm. 
This rate matches the minimax lower bound for regression on
$\mathcal{H}(m,\mathbf{w},t,\beta,K)$, extending
Proposition~\ref{prop:sh} from mean regression with Gaussian errors to quantile
regression with arbitrary error distributions.
\end{proposition}

Propositions~\ref{prop:sh} and \ref{prop:padilla} provide the theoretical foundation for the GenTE architecture. In GenTE, each subnetwork estimates a nonparametric function, $\mu(x, q)$, $\theta(x, q)$, or $\pi(x)$, that maps high-dimensional covariates to a scalar output. When these target functions possess compositional structure (as it is plausible for the quantile function of potential outcomes), sparse deep ReLU networks can achieve convergence rates that depend on the intrinsic complexity of each compositional layer of the oracle function rather than the ambient covariate dimension~$p$.

\subsection{Modeling and Estimation}

We now restrict \eqref{eq:composite} to the model we fit. Taken together, we have the outcome model in \eqref{eq:observed} and the treatment assignment model in \eqref{eq:selection} which involve three unknown
functions: the baseline $\mu(x)$, the treatment effect $\theta(x)$,
and the propensity score $\pi(x) = F_u(\nu(x))$. The first two govern the
outcome process and the third the treatment assignment process. Under overlap and conditional unconfoundedness
all three are nonparametrically identified. When treatment is well predicted by the covariates, however, $\mu(\cdot)$ and
$\theta(\cdot)$ account for overlapping variation in the observed outcome, and a
flexible estimator of $\mu$ that is unaware of how treatment is assigned may absorb part
of the treatment effect where $\pi(x)$ is close to zero or one, or leave baseline
variation to be taken up by $\theta(x)$.  As a result, \citet{hahn2020bayesian}
recommends supplying the propensity score as an additional input to the baseline,
\begin{equation}
\mathbb{E}\bigl[Y \mid X = x,\, D = d\bigr]
= \mu\bigl(x,\, \pi(x)\bigr) + \theta(x)\, d ,
\label{eq:hahn}
\end{equation}
with $\mu$ and $\theta$ modeled separately. Because $\pi(x)$ is itself a function of
$x$, \eqref{eq:hahn} is a reparameterization rather than a restriction: the class of
conditional means it can represent is unchanged, and what the additional argument alters
is the complexity the regularizer assigns to functions within that class. The choice of
$\pi(\cdot)$ is likewise not arbitrary. Since $F_u$ is strictly increasing, $\pi(x)$
is a monotone transformation of the selection index $\nu(x)$, and $\pi(x)$ is a
balancing score in the sense of \citet{rosenbaum1983central}, so it condenses the
assignment mechanism into a single coordinate. Making that coordinate available to the
baseline allows $\mu$ to represent outcome variation that moves with the propensity to
be treated and can improve estimation as discussed by \citet{hahn2018regularization, hahn2020bayesian}.

The parameterization in this paper is the quantile-level counterpart of \eqref{eq:hahn},
obtained by letting both the baseline and the treatment effect inherit the compositional
form \eqref{eq:composite}, with the baseline retaining the propensity score embedding as
an additional argument. Let $e_{\pi}\colon\mathbb{R}^{p}\to\mathbb{R}^{8}$ denote the
embedding produced by the propensity network, from which the propensity score $\pi(x)$ is
obtained by a small output network. Equation~\eqref{eq:model} defines the class of quantile
predictors over which we optimize, together with the propensity score model:
\begin{align}\label{eq:model}
Q(q; x, d)
&\;=\; \mu\bigl(x,\, e_{\pi}(x),\, q\bigr) \;+\; \theta(x, q)\, d ,\\
\pi(x) &\;=\; \Pr\bigl(D = 1 \mid X = x\bigr) \approx  \Lambda\bigl(\mathcal{G}_{\pi}(e_{\pi}(x))\bigr) ,\nonumber
\end{align}
where $\mathcal{G}_{\pi}\colon\mathbb{R}^{8}\to\mathbb{R}$ maps the propensity embedding
to a scalar index and $\Lambda(z) = (1+e^{-z})^{-1}$ is the logistic function,
so that the score is recovered from the embedding by a small output network. The two components in the outcome model are built symmetrically from a common quantile embedding,
\begin{equation}
\mu\bigl(x, e_{\pi}(x), q\bigr)
\approx \mathcal{G}_{\mu}\bigl(\psi_{\mu}(x, e_{\pi}(x))\odot\varphi(q)\bigr),
\qquad
\theta(x,q)
\approx \bigl[\mathcal{G}_{\theta}\bigl(\psi_{\theta}(x)\odot\varphi(q)\bigr)\bigr]_{2},
\label{eq:model-components}
\end{equation}
where $\psi_{\mu}$ and $\psi_{\theta}$ are the covariate encoders of the baseline and
effect subnetworks, $\varphi$ is the cosine quantile embedding \eqref{eq:cosine_embed}, and
$\mathcal{G}_{\mu}\colon\mathbb{R}^{k}\to\mathbb{R}$ and
$\mathcal{G}_{\theta}\colon\mathbb{R}^{k}\to\mathbb{R}^{2}$ are output heads. The second
coordinate of $\mathcal{G}_{\theta}$ carries the quantile-indexed effect. \footnote{The algorithm also supports a specification targeting the difference of
conditional medians, $\operatorname{Med}[Y_s(1)\mid X_s = x] -
\operatorname{Med}[Y_s(0)\mid X_s = x]$ in addition to CATE. In that specification, the
first coordinate of the output head is a second outcome prediction fitted by absolute
deviations, which stabilizes the level of the quantile surface. It is inactive in the
specification reported here and enters no estimand. Further details are provided in
\ref{app:loss}.}  Both arms are therefore quantile-specific:
$Q(q;x,0) = \mu(x, e_{\pi}(x), q)$ estimates the control conditional quantile function and
$Q(q;x,1) - Q(q;x,0) = \theta(x,q)$ the quantile treatment effect, so that
$\mathbb{E}_{q}[\theta(x,q)]$ recovers \eqref{eq:cate-identified}. The main
asymmetry between the two subnetworks is that $e_{\pi}(x)$ enters the baseline encoder and not
the effect encoder.  A related specification is the partially linear model of
\citet[equations (1.1) and (1.2)]{chernozhukov2018double}, in which the
treatment enters with a constant coefficient and the covariates through an
unrestricted nuisance function. Our system relaxes the homogeneity of treatment effects
by letting the coefficient on $D_s$ vary with the covariates and with the
quantile level, and models the assignment mechanism simultaneously.

The architecture in \eqref{eq:model} represents the main methodological
contribution of the paper. The compositional quantile parameterization builds
on the Implicit Quantile Network of \citet{dabney2018implicit}, which was
introduced for predictive distributional learning. Here, however, the
quantile-indexed networks parameterize latent components of a causal model
rather than only the distribution of an observed outcome. In particular, the
baseline function and the conditional quantile treatment-effect surface are
estimated jointly with the treatment-assignment mechanism in an observational
setting in which treatment may depend on observed covariates. Thus, while the
underlying compositional quantile-covariate representation is adapted from \cite{dabney2018implicit}, its use
to jointly parameterize and estimate the baseline, conditional quantile treatment
effect, and the propensity score is specific to the proposed GenTE
architecture.

Two features of
\eqref{eq:model} depart from \eqref{eq:hahn}. First, the baseline is supplied the
propensity \emph{embedding} rather than the scalar score. At the population level, if the propensity score is exactly recovered from
$e_{\pi}(x)$, then $e_{\pi}(x)$ is itself a balancing score in the sense of
\citet{rosenbaum1983central}, and a finer one. Hence, it retains
the property that motivates \eqref{eq:hahn} while giving the baseline
encoder a richer summary of the assignment mechanism. The learned embedding in the network is designed to approximate such a representation.  Second, $\pi(x)$ is not replaced by
a first-stage estimate: the propensity, baseline, and effect subnetworks are trained
simultaneously under a single loss function (described in \ref{app:loss}), so gradients from the outcome reach $e_{\pi}$ as
well.  \autoref{fig:architecture} schematically summarizes the algorithm.

Once estimated, the effect function delivers a quantile-specific treatment effect
directly: a single evaluation at any level $q \in (0,1)$ returns $\hat\theta(x,q)$, with
no need to fit and difference separate models for the treated and control arms. The CATE is recovered as an integrated functional of the estimated quantile treatment effect surface rather than requiring a separate fit. We evaluate
the integral in two ways. The algorithm uses Monte Carlo integration,
averaging over $M = 500$ levels $q_{m}$ drawn randomly from
$\mathrm{Unif}(0,1)$,
\begin{equation}
\hat{\theta}(x) \;=\; \frac{1}{M}\sum_{m=1}^{M}\hat{\theta}\!\left(x,\, q_{m}\right),
\label{eq:cate_est}
\end{equation}
with each of the $3$ ensemble members evaluated at its own set of draws. The
alternative replaces the draws by a fixed grid of $19$ equally spaced levels, which is
also the grid on which quantile treatment effects are reported when the curve
$\hat\theta(x,\cdot)$ itself is of interest. The two are quadratures of the same
integral and differ in no other respect: \autoref{tab:gente_sep_joint} of
\ref{app_T_learner} reports both, and the resulting mean squared errors differ by less
than one percent in every cell, so the estimates do not depend on how the
integral is approximated.

\begin{figure}[H]
\centering
\resizebox{\textwidth}{!}{%
\begin{tikzpicture}[
    every node/.style={font=\normalsize},
    inp/.style={circle,draw,thick,minimum size=0.95cm,fill=blue!12},
    enc/.style={rectangle,draw,thick,rounded corners=2pt,
                minimum height=1.8cm,minimum width=1.3cm},
    encpi/.style={enc,fill=orange!22},
    encmu/.style={enc,fill=teal!25},
    encth/.style={enc,fill=purple!20},
    embed/.style={rectangle,draw,thick,rounded corners=2pt,
                  minimum height=0.9cm,minimum width=1.8cm,fill=orange!12},
    featmu/.style={rectangle,draw,thick,rounded corners=2pt,
                   minimum height=0.9cm,minimum width=1.8cm,fill=teal!12},
    featth/.style={rectangle,draw,thick,rounded corners=2pt,
                   minimum height=0.9cm,minimum width=1.8cm,fill=purple!12},
    cosb/.style={rectangle,draw,thick,rounded corners=2pt,
                 minimum height=0.9cm,minimum width=2.0cm,fill=yellow!25},
    gmubox/.style={rectangle,draw,thick,rounded corners=2pt,
                   minimum height=1.0cm,minimum width=1.9cm,fill=teal!12},
    gthbox/.style={rectangle,draw,thick,rounded corners=2pt,
                   minimum height=1.0cm,minimum width=1.9cm,fill=purple!12},
    prod/.style={circle,draw,thick,minimum size=0.8cm,fill=gray!15},
    piout/.style={circle,draw,thick,minimum size=0.9cm,fill=orange!30},
    muout/.style={circle,draw,thick,minimum size=1.1cm,fill=teal!30},
    thout/.style={circle,draw,thick,minimum size=1.1cm,fill=purple!18},
    outbox/.style={rectangle,draw,thick,rounded corners=3pt,
                   minimum height=1.1cm,minimum width=6.4cm,fill=teal!15},
    catebox/.style={rectangle,draw,thick,rounded corners=3pt,
                    minimum height=1.1cm,minimum width=6.4cm,fill=purple!15},
    arrow/.style={-{Stealth[length=2.6mm]},thick},
    darrow/.style={-{Stealth[length=2.6mm]},thick,dashed,gray!70!black},
    sect/.style={font=\small,text width=3.6cm,align=center},
    dim/.style={font=\footnotesize,gray!50!black}
]
\node[sect] at (-6.6,3.2) {Inputs};
\node[inp] (x) at (-6.6,2.0) {$X_s$};
\node[inp] (d) at (-6.6,0.6) {$D_s$};
\node[inp] (q) at (-6.6,-0.8) {$q$};

\node[sect] at (-2.0,6.0) {Propensity network};
\node[encpi] (hpi) at (-2.0,4.6) {$h^{\pi}$};
\node[embed] (epi) at (0.6,4.6) {$e_{\pi}(X_s)$};
\node[dim] at (2.4,4.6) {$\in\mathbb{R}^{8}$};
\node[piout] (pihat) at (0.6,6.6) {$\hat\pi(X_s)$};

\node[sect] at (-2.0,3.5) {Baseline network};
\node[encmu] (hmu) at (-2.0,2.0) {$h^{\mu}$};
\node[featmu] (psimu) at (2.6,2.0) {$\psi_{\mu}(X_s, e_{\pi}(X_s))$};
\node[dim] at (2.6,1.2) {$\in\mathbb{R}^{64}$};

\node[sect] at (-2.0,0.2) {Quantile embedding};
\node[cosb] (cos) at (-2.0,-0.8) {$\cos(\pi i q)$};
\node[cosb] (phi) at (2.6,-0.8) {$\varphi(q)$};
\node[dim] at (2.6,-1.6) {$\in\mathbb{R}^{64}$};

\node[sect] at (-2.0,-2.2) {Treatment-effect network};
\node[encth] (hth) at (-2.0,-3.6) {$h^{\theta}$};
\node[featth] (psith) at (2.6,-3.6) {$\psi_{\theta}(X_s)$};
\node[dim] at (2.6,-4.4) {$\in\mathbb{R}^{64}$};

\node[sect] at (7.8,3.3) {Quantile interaction};
\node[prod] (odotmu) at (5.4,2.0) {$\odot$};
\node[gmubox] (gmu) at (7.8,2.0) {$\mathcal{G}_{\mu}(\cdot)$};
\node[muout] (muhat) at (10.6,2.0) {$\hat\mu(X_s,q)$};

\node[prod] (odotth) at (5.4,-3.6) {$\odot$};
\node[gthbox] (gth) at (7.8,-3.6) {$\mathcal{G}_{\theta}(\cdot)$};
\node[thout] (that) at (10.6,-3.6) {$\hat\theta(X_s,q)$};

\node[sect] at (16.4,3.3) {Outputs};
\node[outbox] (y0) at (16.4,2.0)
  {$\hat F^{-1}_{Y\mid X_s,D=0}(q) = \hat\mu(X_s,q)$};
\node[outbox] (y1) at (16.4,-0.8)
  {$\hat F^{-1}_{Y\mid X_s,D=1}(q) = \hat\mu(X_s, q) + \hat\theta(X_s,q)$};
\node[catebox] (cate) at (16.4,-5.8)
  {$\hat\theta(X_s) = \dfrac{1}{M}\displaystyle\sum_{m=1}^{M}\hat\theta(X_s,q_m)$};

\draw[thick] (x) -- (-4.6,2.0);
\draw[thick] (-4.6,4.6) -- (-4.6,-3.6);
\draw[arrow] (-4.6,4.6) -- (hpi);
\draw[arrow] (-4.6,2.0) -- (hmu);
\draw[arrow] (-4.6,-3.6) -- (hth);

\draw[arrow] (hpi) -- (epi);
\draw[arrow] (epi) -- (pihat);
\draw[darrow] (d) -- (-8.2,0.6) -- (-8.2,6.6)
  -- node[midway,above,text=black]{estimation target} (pihat);

\draw[arrow] (epi) -- (0.6,3.3) -- (-2.0,3.3) -- (hmu.north);

\draw[arrow] (q) -- (cos);
\draw[arrow] (cos) -- (phi);
\draw[arrow] (phi.north east) -- (odotmu.south west);
\draw[arrow] (phi.south east) -- (odotth.north west);

\draw[arrow] (hmu) -- (psimu);
\draw[arrow] (psimu) -- (odotmu);
\draw[arrow] (odotmu) -- (gmu);
\draw[arrow] (gmu) -- (muhat);

\draw[arrow] (hth) -- (psith);
\draw[arrow] (psith) -- (odotth);
\draw[arrow] (odotth) -- (gth);
\draw[arrow] (gth) -- (that);

\draw[arrow] (muhat) -- (y0.west);
\draw[arrow] (muhat) -- (12.4,2.0) -- (12.4,-0.8) -- (y1.west);
\draw[arrow] (that) -| (y1.south);
\draw[arrow] (that.south) -- (10.6,-5.8) -- (cate.west);
\end{tikzpicture}%
}
\caption{Architecture of the generative causal learner (GenTE). Three subnetworks with
separate covariate encoders are estimated jointly. The propensity network produces an
$8$-dimensional embedding $e_{\pi}(X_s)$, from which the propensity score $\hat\pi(X_s)$
is obtained and which is also passed to the baseline encoder as a second argument,
implementing the reparameterization in \eqref{eq:model}. Both arms are quantile-specific
and built from the same compositional form: the baseline network returns covariate features
$\psi_{\mu}(X_s, e_\pi(X_s))$ and the treatment-effect network returns $\psi_{\theta}(X_s)$, each
composed with the same cosine quantile embedding $\varphi(q)$ by element-wise
multiplication and mapped through an output head, $\mathcal{G}_{\mu}$ and
$\mathcal{G}_{\theta}$, to $\hat\mu(X_s,q)$ and $\hat\theta(X_s,q)$. The control arm is
the baseline evaluated at $q$ and the treated arm adds the effect at the same $q$, so
$\hat\mu(\cdot,q)$ estimates the control conditional quantile function and
$\hat\theta(\cdot,q)$ the quantile treatment effect. The CATE is the average of the quantile-specific effects, computed either over
$M = 500$ levels drawn from $\mathrm{Unif}(0,1)$ or over a fixed grid of $19$
levels (as in \autoref{tab_mse}), both numerical approximations to
$\int_{0}^{1}\hat\theta(X_s,q)\,dq$. All hidden layers use ReLU activations and have width
$16$: two in the propensity path, four in the baseline, and three in the effect module.
The covariate encoders and the quantile embedding have dimension $64$, the propensity
embedding dimension $8$, and the cosine basis uses $32$ frequencies $\cos(\pi i q)$,
$i = 0,\dots,31$. Network weights are optimized by adaptive moment estimation (Adam).
The composite loss function is described in \ref{app:loss}, and the algorithm is
implemented in the Python package
\href{https://github.com/marnare/generativeAI/blob/main/src/GBCcausal.py}{\texttt{GBCcausal/GenTE.py}}. }
\label{fig:architecture}
\end{figure}

The output network of the treatment effect subnetwork, the baseline, and the propensity
network are obtained jointly under a single composite loss function described by \ref{app:loss}. At each iteration two quantile levels $q$ and $q'$ are drawn independently from
$\mathrm{Unif}(0,1)$ and the network is evaluated at both, so that the
optimization targets minimizing the loss $\mathbb{E}_{q,q'}[\mathcal L(w; q, q')]$ over the network
parameters $w$. The loss function has three active terms. The check function of
quantile regression, evaluated at $q$, is minimized when the fitted quantile
equals the conditional $q$-quantile of the observed outcome given $(X_s, D_s)$,
and is the term that identifies the quantile structure. It carries the largest
weight in the proposed simulations and the application among the outcome terms. A binary cross-entropy on the treatment indicator models the treatment assignment mechanism, and is what makes the embedding $e_{\pi}(X_s)$ passed to the baseline informative about it. A penalty on quantile crossings compares the fits at $q$
and $q'$ and is positive exactly when the estimate at the lower level exceeds
the estimate at the higher one, which a conditional quantile function cannot do
by construction. Because the levels are redrawn
at every iteration, one network is trained across a continuum of quantile levels
rather than a fixed grid, and $\widehat\theta(X_s, q)$ is available at every
$q \in (0,1)$ after a single training run. The exact form of each term, the
weights attached to them, and a fourth term used only when the target is a
conditional median effect are given in \ref{app:loss}.

An alternative is to model \eqref{eq_cate_composition} separately in each treatment arm \(d\in \{0, 1\}\) and subtract,
$\hat\theta(x) = \int_0^1 \bigl[\hat{g}_1(\cdot) - \hat{g}_0(\cdot)\bigr] dq$. \footnote{This approach is known as a T-learner, and GenTE implemented at  \href{https://github.com/marnare/generativeAI/blob/main/src/GBCcausal.py}{\texttt{GBCcausal/GenTE.py}} does support such an estimator under \texttt{estimate\_qte\_separate}. } We instead jointly estimate the latent representations, baseline outcome component, conditional average treatment effect, and the propensity model for three main reasons. First, the treatment effect is parameterized explicitly rather than recovered as the difference between two separately fitted response surfaces. This permits the complexity of the treatment-effect function to be regularized directly, potentially reducing spurious heterogeneity when the treatment effect is small relative to baseline variation. Under differencing the two arm-specific biases need not cancel while estimation uncertainty from both surfaces contributes to the uncertainty of their difference, which is costly when the
effect is small relative to the variation in the baseline. Second, both treatment arms inform the
baseline, and this matters exactly when treatment is well predicted by the covariates:
where $\pi(x)$ is close to zero or one the thinner arm contains few observations and,
fitted on its own subsample, must extrapolate without borrowing strength from the other.
Third, the propensity embedding is learned jointly with the outcome components and supplied directly to the baseline network. Although separately fitted treatment-specific models may also incorporate estimated propensity scores, GenTE differs by allowing the representation of the assignment mechanism to be updated simultaneously with the baseline and treatment-effect functions through a shared optimization objective. Monte Carlo experiments in \autoref{tab:gente_sep_joint} of  \ref{app_T_learner} compare the joint estimator with a T-learner that fits the treated and control quantile functions separately. The joint specification attains a lower mean squared error for the CATE than the separate-arm estimator in almost every design: the average reduction exceeds 50\%. In the additive designs with larger samples the mean squared error falls by an order of magnitude, and the Monte Carlo standard deviation of the error declines in parallel.

\section{Monte Carlo Simulations} 
\label{sec:simulation}

\subsection{Functional Forms of Conditional Average Treatment Effects}\label{sec_sims_mse}

We evaluate the performance of the proposed generative learner using simulated data for which the true
conditional average treatment effect is known. The data-generating process follows a standard
structural equation model as defined by \eqref{eq:observed}: \(
Y_s = \mu(X_s) + \theta(X_s)\,D_s + \varepsilon_s\).
Here, $X_s\in\mathbb{R}^p$ is a vector of covariates, $D_s\in\{0,1\}$ is a binary treatment indicator, $Y_s$ is a continuous outcome,
$\theta(X_s)$ is a conditional average treatment effect, and $\varepsilon_s$ is an idiosyncratic error term. For each simulated sample, we draw $S$ independent observations of covariates \(
X_s \sim \mathcal N(0,I_p),
\) with varying covariate dimension \(p\). The baseline outcome function is linear in covariates, \(
\mu(X_s) = X_s^\top \beta_\mu,
\) where $\beta_\mu \in \mathbb{R}^p$ is drawn once per simulation from $\mathcal N(0,0.5^2 I_p)$.
This specification generates moderate baseline heterogeneity.

We consider several designs for the conditional treatment effect $\theta(X_s)$ in order to evaluate the
ability of the learner to recover different forms of heterogeneity:
\begin{align*}
\text{Linear:} \qquad
&\theta(X_s) = 1 + X_{s1} + 0.5 X_{s2}, \\
\text{Nonlinear:} \qquad
&\theta(X_s) = 1 + X_{s1}^2 + 0.5 X_{s2}^2, \\
\text{Interaction:} \qquad
&\theta(X_s) = 1 + X_{s1} X_{s2}, \\
\text{Nonlinear interaction:} \qquad
&\theta(X_s) = 1 + X_{s1}^2 X_{s2}.
\end{align*}

The treatment assignment mechanism in each design depends on covariates through a logistic model:
\[
\mathbb{P}(D_s=1 \mid X_s)
=
\Lambda \bigl(0.3 X_{s1} - 0.2 X_{s2} + 0.1 X_{s3}\bigr),
\]
where $\Lambda(\cdot)$ denotes the logistic function in line with  Subsection \ref{sec_treat_assign} (when \(p = 2\), the assignment mechanism is defined as \(\Lambda \bigl(0.3 X_{s1} - 0.2 X_{s2}\bigr)\)). This induces selection on observable covariates and creates
confounding between $D_s$ and $Y_s$ through $X_s$.  The outcome disturbance is drawn independently as
\(
\varepsilon_s \sim \mathcal N(0,0.5^2)
\). \autoref{fig:assignment_mse} in \ref{app_assignment} shows that with the CATE held fixed, the performance of GenTE is robust to alternative treatment assignment mechanisms, such as the randomized, linear logit, a
nonlinear logit, or a trigonometric propensity. \autoref{fig:snr_mse} additionally measures the performance across various levels of the signal-to-noise ratio. Because the oracle function $\theta(X_s)$ is known by construction, the simulation design allows us to
evaluate the precision of estimated CATEs directly. For each algorithm, it is measured by the mean squared error (MSE) of the
estimated effect against the simulated truth on the held-out half \(S_{\text{test}}\) of each sample:
\[
\mathrm{MSE} = \frac{1}{S_{\mathrm{test}}}\sum_{s}\bigl(\hat\theta(X_s) -
\theta(X_s)\bigr)^2,
\] 
where \(S_{\mathrm{test}}\) is the number of observations in the test data set, and the summation is taken over the observations in the test fold. The covariates are standardized before estimation, with the location and scale
computed on the training data alone and applied to the test data, so that
no information from the held-out sample enters the fit. Each replication splits
the sample in half, stratified by treatment status: one half is used for
training and the other, unused in estimation, for evaluation.  \footnote{GenTE is trained
for 500 iterations with a learning rate of $0.01$, a weight decay of $10^{-3}$,
and an ensemble of $3$ members by default. The competing estimators are used as supplied
by their implementations: the generalized random forest at its package
defaults, GANITE for 500 iterations, and DML-XGB with gradient-boosted
regression for the outcome and a random forest for the treatment. All methods are run at their package default settings, and no hyperparameter search is
carried out for any estimator, so that none benefits from a tuning effort not extended
to the others. Defaults are not neutral in principle, since they may suit some
estimators better than others on a given design. Hence, the comparison should therefore be read
as indicative of out-of-the-box performance rather than of the best accuracy each method
could attain under tuning.}

\begin{table}[H]
\centering
\caption{Mean squared error (MSE) of the generalized random forest (GRF), double machine learning based on XGBoost (DML-XGB), GANITE, and GenTE for different effect types, sample sizes $S$, and covariate dimensions $p$. Each cell reports the mean MSE with standard deviation in parentheses based on 100 Monte Carlo experiments. \bt{Dark blue} marks the cells in which a benchmark attains a lower MSE than GenTE. The last three columns report the average percent gain of GenTE relative to (i) GRF, (ii) DML-XGB, and (iii) GANITE, averaged over $p\in\{2,4,8\}$ in each row.}
\resizebox{\textwidth}{!}{%
\begin{tabular}{lrrrrrrrrrrrrrrr}
\toprule
& \multicolumn{3}{c}{GRF} & \multicolumn{3}{c}{DML-XGB} & \multicolumn{3}{c}{GANITE} & \multicolumn{3}{c}{GenTE} & \multicolumn{3}{c}{Avg.\ \% gain (GenTE)} \\
\cmidrule(lr){2-4}\cmidrule(lr){5-7}\cmidrule(lr){8-10}\cmidrule(lr){11-13}\cmidrule(lr){14-16}
Sample size
& $p{=}2$ & $p{=}4$ & $p{=}8$
& $p{=}2$ & $p{=}4$ & $p{=}8$
& $p{=}2$ & $p{=}4$ & $p{=}8$
& $p{=}2$ & $p{=}4$ & $p{=}8$
& vs GRF & vs DML & vs GANITE \\
\midrule
\multicolumn{16}{l}{Linear} \\
\midrule
200 & 0.800 & 0.861 & 0.994 & 0.465 & 0.610 & 0.808 & 1.029 & 0.874 & 1.123 & 0.107 & 0.345 & 0.712 & 58.3\% & 44.1\% & 62.3\% \\
  & (0.188) & (0.188) & (0.222) & (0.148) & (0.166) & (0.215) & (0.362) & (0.294) & (0.381) & (0.067) & (0.165) & (0.380) &  &  &  \\
500 & 0.298 & 0.354 & 0.500 & 0.225 & 0.281 & 0.419 & 0.987 & 0.889 & 1.118 & 0.064 & 0.222 & 0.409 & 44.7\% & 31.7\% & 77.3\% \\
  & (0.063) & (0.067) & (0.130) & (0.070) & (0.067) & (0.116) & (0.287) & (0.330) & (0.344) & (0.046) & (0.138) & (0.232) &  &  &  \\
1000 & 0.161 & 0.203 & \bt{0.293} & 0.126 & \bt{0.161} & \bt{0.242} & 1.044 & 0.855 & 1.127 & 0.053 & 0.170 & 0.302 & 26.7\% & 9.2\% & 82.7\% \\
  & (0.034) & (0.039) & (0.071) & (0.033) & (0.037) & (0.058) & (0.374) & (0.272) & (0.320) & (0.046) & (0.101) & (0.147) &  &  &  \\
\addlinespace
\midrule
\multicolumn{16}{l}{Nonlinear} \\
\midrule
200 & 2.646 & 2.529 & 2.525 & 2.381 & 2.372 & 2.455 & 5.097 & 4.952 & 4.995 & 0.346 & 0.660 & 1.457 & 67.7\% & 66.1\% & 83.6\% \\
  & (0.943) & (0.701) & (0.766) & (0.953) & (0.664) & (0.768) & (1.293) & (1.042) & (1.152) & (0.297) & (0.393) & (0.633) &  &  &  \\
500 & 1.712 & 1.714 & 2.149 & 1.365 & 1.364 & 1.770 & 4.926 & 4.765 & 5.030 & 0.174 & 0.348 & 0.688 & 79.2\% & 74.3\% & 91.8\% \\
  & (0.477) & (0.443) & (0.533) & (0.459) & (0.459) & (0.522) & (0.776) & (0.663) & (0.756) & (0.100) & (0.231) & (0.305) &  &  &  \\
1000 & 1.026 & 1.073 & 1.377 & 0.825 & 0.829 & 1.051 & 4.877 & 4.893 & 5.103 & 0.137 & 0.275 & 0.565 & 73.3\% & 65.5\% & 93.5\% \\
  & (0.295) & (0.273) & (0.376) & (0.266) & (0.236) & (0.319) & (0.557) & (0.531) & (0.569) & (0.063) & (0.119) & (0.259) &  &  &  \\
\addlinespace
\midrule
\multicolumn{16}{l}{Interaction} \\
\midrule
200 & 1.079 & 1.002 & \bt{1.024} & 1.016 & 0.993 & \bt{1.017} & 1.123 & 1.189 & 1.526 & 0.358 & 0.578 & 1.169 & 31.7\% & 30.5\% & 47.6\% \\
  & (0.346) & (0.270) & (0.271) & (0.358) & (0.270) & (0.287) & (0.360) & (0.302) & (0.365) & (0.206) & (0.258) & (0.453) &  &  &  \\
500 & 0.785 & 0.880 & 1.009 & 0.750 & 0.836 & 0.972 & 1.089 & 1.199 & 1.626 & 0.227 & 0.377 & 0.666 & 54.1\% & 52.1\% & 68.9\% \\
  & (0.162) & (0.174) & (0.182) & (0.167) & (0.172) & (0.186) & (0.254) & (0.246) & (0.295) & (0.108) & (0.142) & (0.276) &  &  &  \\
1000 & 0.607 & 0.753 & 0.907 & 0.561 & 0.678 & 0.864 & 1.079 & 1.204 & 1.623 & 0.203 & 0.308 & 0.521 & 56.1\% & 52.7\% & 74.5\% \\
  & (0.119) & (0.102) & (0.118) & (0.122) & (0.101) & (0.115) & (0.184) & (0.191) & (0.244) & (0.094) & (0.115) & (0.191) &  &  &  \\
\addlinespace
\midrule
\multicolumn{16}{l}{Nonlinear interaction} \\
\midrule
200 & 2.882 & 2.470 & \bt{2.663} & 2.756 & 2.408 & \bt{2.591} & 3.105 & 2.746 & \bt{2.996} & 1.954 & 2.155 & 3.197 & 8.3\% & 5.4\% & 17.3\% \\
  & (2.093) & (1.407) & (1.422) & (2.023) & (1.385) & (1.414) & (2.053) & (1.326) & (1.460) & (1.640) & (1.139) & (1.550) &  &  &  \\
500 & 2.225 & 2.029 & \bt{2.429} & 2.290 & 2.074 & \bt{2.483} & 2.804 & 2.709 & 3.220 & 1.403 & 1.632 & 2.519 & 17.6\% & 19.5\% & 37.2\% \\
  & (0.938) & (0.807) & (0.886) & (0.941) & (0.803) & (0.872) & (1.027) & (0.905) & (1.004) & (0.930) & (0.787) & (1.019) &  &  &  \\
1000 & 1.941 & 1.853 & \bt{2.158} & 1.952 & 1.807 & \bt{2.141} & 2.830 & 2.774 & 3.235 & 1.364 & 1.525 & 2.231 & 14.7\% & 13.9\% & 42.6\% \\
  & (0.670) & (0.579) & (0.565) & (0.680) & (0.593) & (0.594) & (0.771) & (0.690) & (0.693) & (0.641) & (0.641) & (0.602) &  &  &  \\
\bottomrule
\end{tabular}%
}
\label{tab_mse}
\end{table}

\autoref{tab_mse} reports out-of-sample mean squared error, averaged
over 100 Monte Carlo experiments.  GenTE attains the lowest error of the
four estimators in 30 of the 36 design cells, and the row-wise average
reduction relative to a given benchmark ranges from 5.4\% to 93.5\%.
The margin is widest in the nonlinear design, where GenTE leads in
every cell and lowers the error by 65\% to 94\%, and narrowest in the
nonlinear interaction design, in which all four estimators have
comparatively large errors.  The six cells where GenTE does not lead,
marked in dark blue, are concentrated at the largest covariate
dimension: five arise at $p = 8$, three of them in the nonlinear
interaction design.

\begin{figure}[H]
\centering

\begin{subfigure}[t]{0.48\textwidth}
  \centering
  \includegraphics[width=\linewidth]{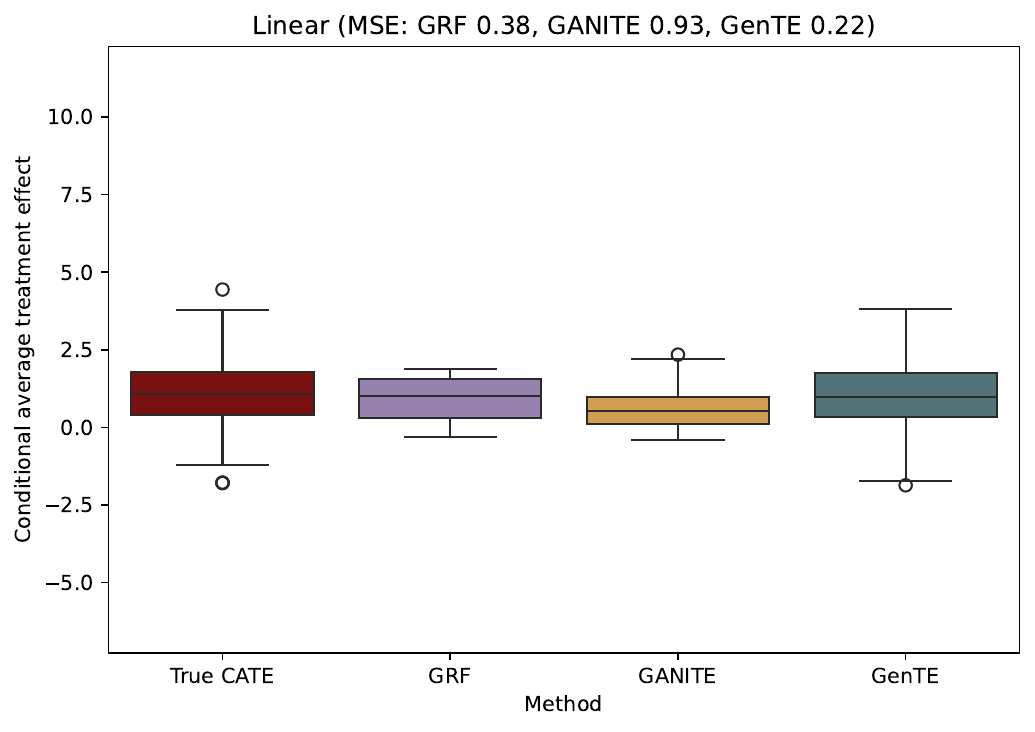}
  \caption{Linear}
\end{subfigure}\hfill
\begin{subfigure}[t]{0.48\textwidth}
  \centering
  \includegraphics[width=\linewidth]{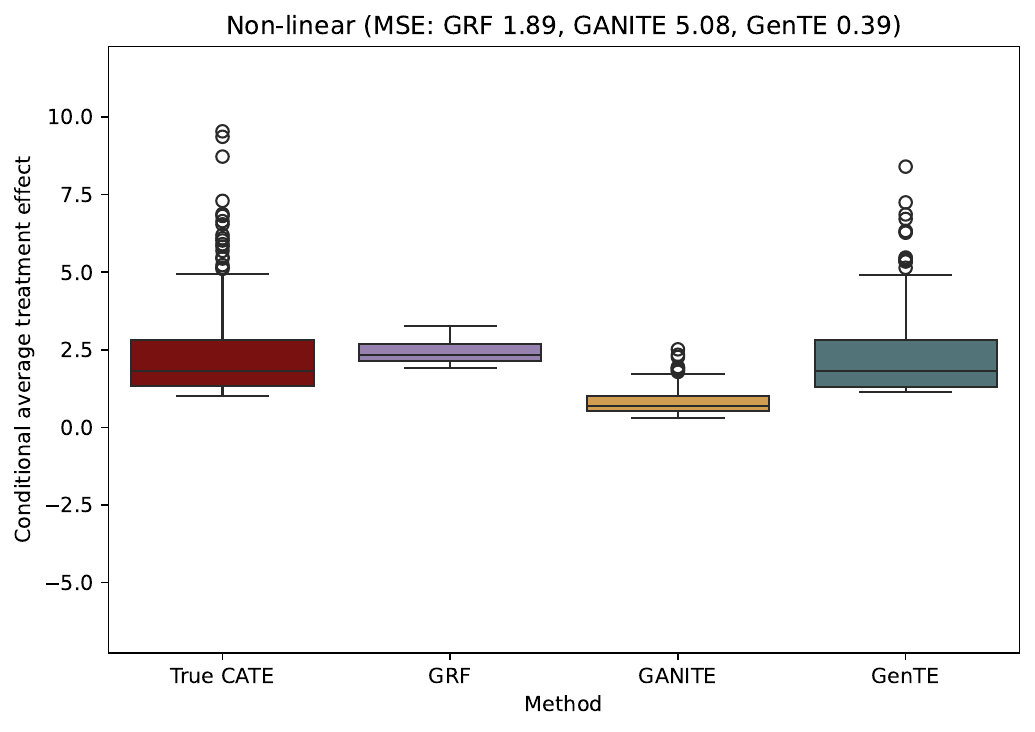}
  \caption{Nonlinear}
\end{subfigure}

\vspace{0.6em}

\begin{subfigure}[t]{0.48\textwidth}
  \centering
  \includegraphics[width=\linewidth]{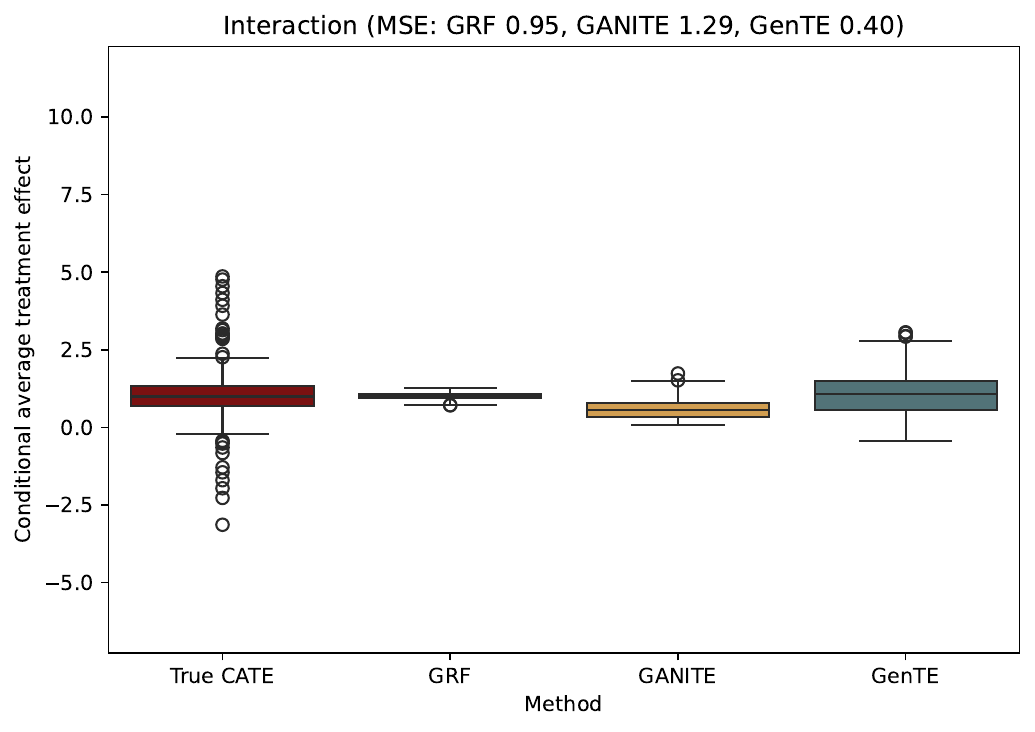}
  \caption{Interaction}
\end{subfigure}\hfill
\begin{subfigure}[t]{0.48\textwidth}
  \centering
  \includegraphics[width=\linewidth]{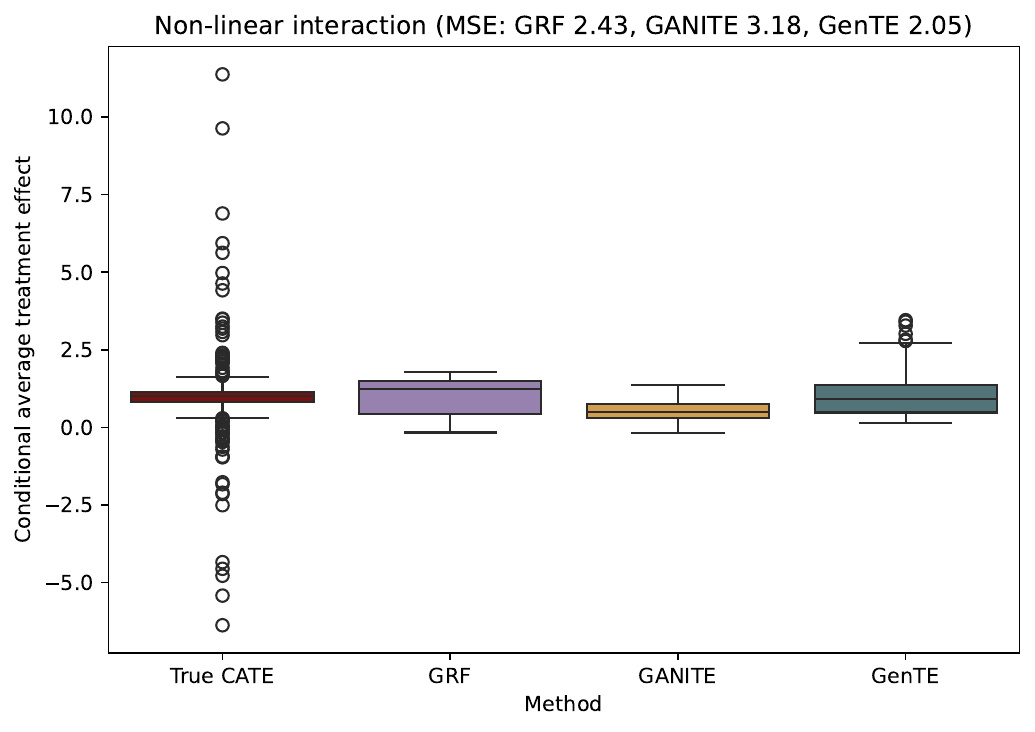}
  \caption{Nonlinear interaction}
\end{subfigure}

\caption{Empirical distributions of the conditional average treatment effect (CATE) estimates, evaluated on the held-out test set and averaged over 100 Monte Carlo replications. GenTE denotes the generative treatment effect learner proposed in this article. GRF denotes the generalized random forest estimator \citep{athey2019generalized}. GANITE represents generative adversarial networks for individualized treatment effects \citep{yoon2018ganite}. True CATE is the simulated oracle conditional average treatment effect. Each simulation uses 500 observations and four covariates. }
\label{fig:sim_comparison_2x2}
\end{figure}

Additionally, \autoref{fig:sim_comparison_2x2} and \autoref{fig:sim_comparison_ganite_scatter} summarize the performance of the proposed generative learner (GenTE)
relative to benchmark methods. Each panel in  \autoref{fig:sim_comparison_2x2} reports the
distribution of conditional average treatment effects, estimated on held-out test data, together with the true CATE distribution, averaged over 100 simulated experiments. GenTE substantially outperforms benchmark models in terms of out-of-sample mean squared error. In the linear case in panel (a), all methods recover the center of the oracle CATE distribution well,
but GenTE has the entire distribution markedly closer to the truth. This gap widens further in the nonlinear design, where GRF struggles to estimate the
right tail of the CATE distribution, while GenTE closely recovers the oracle distribution and achieves
substantial reduction in MSE.

The advantages of the proposed generative learner are also evident in  \autoref{fig:sim_comparison_ganite_scatter} of \ref{app_scatters}.  GenTE’s predictions lie in a narrow band along the perfect-fit line, with low spread, and a shape closely following the diagonal. GRF is second-best: its points follow the diagonal but with more scatter and increasing bias with nonlinear and interaction effects, especially for higher levels of true CATEs. GANITE shows the worst performance: predictions are more dispersed and often biased, with substantial deviations from the diagonal and, in nonlinear settings, a clear tendency to compress high true CATEs toward zero. Spread and bias worsen with complexity for all methods, but GenTE maintains an accurate approximation of the true effects and most accurate location across designs, while GANITE’s location and shape deteriorate the most when effects are nonlinear functions of covariates.

\ref{app_error_distributions} in addition compares two parameterizations of the same
estimand. In the first, the treatment effect is a function of the covariates
alone, $\theta(x)$, and is fitted by squared loss. In the second, it is indexed
by the quantile level, $\theta(x,q)$, is fitted by the composite loss of
\ref{app:loss}, and is aggregated over $q$ to recover $\theta(x)$. Width, depth,
the propensity embedding, joint estimation and ensemble size are held fixed
across the two, so that the comparison speaks to the parameterization rather
than to other differences between the estimators. Moreover, the conditional law of
the disturbances is varied while the effect function and the assignment
mechanism are held at the specification of Subsection~\ref{sec_sims_mse}. According to \autoref{tab:gente_quantile_direct}, neither parameterization dominates. Where the disturbances are
symmetric and unimodal, the direct specification attains the lower mean squared
error, and tail thickness does not alter this: the ranking under $t-$distributed error terms
matches the Gaussian case. Under a skewed treated arm the two specifications split the cells roughly
evenly, and the quantile-indexed specification is clearly advantageous under
the bimodal mixture design. The comparison concerns the
conditional average effect alone, since the direct specification returns \(\theta(x)\)
and $\theta(x,q)$ is unavailable from it at any level.

Broadly, these findings point to a wider role for the neural network as a tool for statistical estimation and structural learning, rather than prediction alone. In particular, they suggest that flexible neural networks can recover latent quantities which are interpretable and  generalizable even when the corresponding truth is not directly observed.  The proposed architecture embeds the identification structure of the model, so that the fitted components inherit the causal interpretation assigned to them by that structure.  The findings indicate that the quantile-indexed parameterization recovers nonlinear effect functions across a range of error distributions, and is the preferable choice where the conditional outcome distribution is asymmetric or multimodal, or where the quantile-indexed effect is itself of interest.

\subsection{Covariate and Quantile Heterogeneity}\label{app_regimes}

The designs of Subsection~\ref{sec_sims_mse} vary the treatment effect across covariates while holding its distribution across quantile levels fixed. In each of them the treated and control
disturbances share a distribution, so the quantile index carries no information
about the effect and $\theta(x,q) = \theta(x)$ for every $q$. Here, we separate the two channels through which treatment effects can vary, across covariates and across the outcome distribution, by constructing three designs: one without heterogeneity and two in which each channel is active separately.

All three retain the outcome equation \eqref{eq:observed}, the covariate
distribution, the baseline, and the
linear-logit treatment assignment of Section~\ref{sec:simulation}.
They differ only in the effect function and in the scale of the disturbances,
which we now allow to depend on the treatment state,
\begin{equation}
\varepsilon_s(d) \mid X_s = x \sim \mathcal N\bigl(0, \sigma_d^{2}\bigr),
\qquad d \in \{0,1\},
\label{eq:regime-noise}
\end{equation}
with $\sigma_0 = 0.5$ throughout. Because the potential outcomes are then
Gaussian with a state-specific scale, the conditional quantile functions are
available in closed form,
\begin{equation}
F^{-1}_{Y(d)\mid X = x}(q)
= \mu(x) + \theta(x) d + \sigma_d\,\Phi^{-1}(q),
\end{equation}
where $\Phi$ is the standard normal distribution function, so that the
quantile treatment effect of Section~\ref{sec_gente} is
\begin{equation}
\theta(x,q)
= F^{-1}_{Y(1)\mid X = x}(q) - F^{-1}_{Y(0)\mid X = x}(q)
= \theta(x) + (\sigma_1 - \sigma_0)\,\Phi^{-1}(q).
\label{eq:regime-qte}
\end{equation}
The two terms of \eqref{eq:regime-qte} are the two channels: $\theta(x)$ moves
the curve vertically as the covariate profile changes, and
$(\sigma_1 - \sigma_0)\Phi^{-1}(q)$ tilts it as the quantile level changes.
Setting one or the other to a constant isolates a channel:
\begin{align}\label{eq_three_designs}
\text{(a) No heterogeneity:} \quad
&\theta(x) = 1, \;\; \sigma_1 = \sigma_0,
&&\theta(x,q) = 1, \nonumber \\
\text{(b) Covariate heterogeneity:} \quad
&\theta(x) = 1 + x_{1} + 0.5\,x_{2}, \;\; \sigma_1 = \sigma_0,
&&\theta(x,q) = \theta(x), \\
\text{(c) Quantile heterogeneity:} \quad
&\theta(x) = 1, \;\; \sigma_1 = 1.5,
&&\theta(x,q) = 1 + \Phi^{-1}(q).\nonumber 
\end{align}

Design (a) is flat in both arguments, (b) is flat in $q$ and varies in $x$,
and (c) is flat in $x$ and varies in $q$. Since
$\int_0^1 \Phi^{-1}(q)\,dq = 0$, the disturbance scales do not enter the
integral in \eqref{eq:cate-identified}, and the conditional average treatment
effect is
\begin{equation}
\theta(x) = \int_0^1 \theta(x,q)\,dq
= \begin{cases}
1, & \text{in (a) and (c)},\\[2pt]
1 + x_{1} + 0.5\,x_{2}, & \text{in (b)} .
\end{cases}
\end{equation}

Estimation follows Section~\ref{sec:simulation}, with $S = 2000$, $p = 4$, and
$3$ ensemble members trained for 500 iterations at a learning rate of
$0.01$. We report a single sample rather than a Monte Carlo average, since the
object of interest is the shape of $\widehat\theta(x,\cdot)$ rather than its
accuracy in repeated samples, and we set the weight decay to zero. The effect is evaluated at
five fixed covariates, obtained by varying $x_{1}$ over
$[-1.5, 1.5]$ with the remaining coordinates at zero, and at the grid of $19$
quantile levels for reporting quantile treatment effects.

\begin{figure}[H]
\centering
\includegraphics[width=\textwidth]{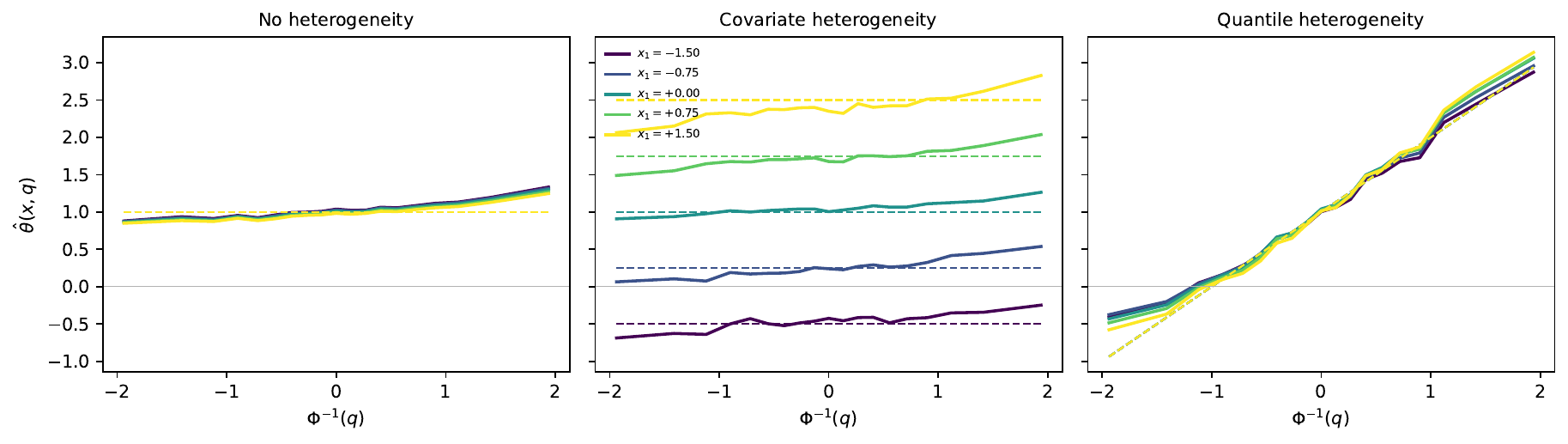}
\caption{Estimated quantile treatment effects $\widehat\theta(x,q)$ (solid)
against the truth (dashed) in the three designs of \eqref{eq_three_designs}, on a
single sample with $S = 2000$ and $p = 4$. Colors index five fixed covariate profiles, obtained by varying $x_{1}$ over $[-1.5, 1.5]$ with the remaining
coordinates held at zero. The same five profiles appear in every panel. The
horizontal axis is $\Phi^{-1}(q)$, on which the truth in the third panel is a
line of slope $\sigma_1 - \sigma_0 = 1$. Vertical separation between colors is
heterogeneity across covariates and slope within a color is heterogeneity
across the outcome distribution. }
\label{fig:qte_regimes}
\end{figure}

\autoref{fig:qte_regimes} reports the estimated effect functions
$\widehat\theta(x,\cdot)$ in the three designs, with the same five covariate
profiles in each panel, so that the two channels of \eqref{eq:regime-qte}
appear as distinct features of the plot: vertical separation between profiles
is heterogeneity across covariates, and slope in $\Phi^{-1}(q)$ is
heterogeneity across the outcome distribution. Under no heterogeneity the
fitted curves coincide across profiles and are close to flat and fixed at one as simulated. Under covariate
heterogeneity they separate in proportion to $\theta(x)$ while remaining flat
in the quantile index, and under quantile heterogeneity they nearly coincide
across profiles while acquiring the slope $\sigma_1 - \sigma_0 = 1$. The
estimator therefore neither imposes quantile variation where the design has
none nor suppresses it where the design has it, with two visible departures:
a mild upward drift at the highest quantile levels in the first two panels,
where the target is flat, and a residual separation across profiles in the
third, where the target does not depend on the covariates. 

The
comparison between the first and third panels of \autoref{fig:qte_regimes} is the substantive one: the
conditional average treatment effect equals $1$ in both, since
$\int_0^1 \Phi^{-1}(q)\,dq = 0$. Thus, an estimator targeting only the conditional mean contrast has the same population estimand in both settings and cannot, by itself, distinguish their distributional effects. That distinction is
delivered by \eqref{eq_cate_composition} as a by-product of the same fit used
to obtain the average effect.

\autoref{fig:qte_regimes} highlights two important implications. The effect of a policy intervention, treatment, or change in an underlying physical quantity, such as the variation in stellar temperature considered below, may differ both across covariates and across different parts of the conditional outcome distribution. For example, a job training program may have different effects for low- and high-income households and, within each group, different effects in the lower and upper tails of the earnings distribution. Similarly, a clinical intervention may vary with patient characteristics while also having different effects across the distribution of health outcomes. The proposed method characterizes these two dimensions of heterogeneity jointly, allowing the researcher to study how effects vary simultaneously across covariates and quantile levels, while also distinguishing variation attributable to each dimension separately.

\section{Empirical Application: The Stefan--Boltzmann Law}

The Stefan--Boltzmann law states that the total power radiated by a black body is
proportional to the fourth power of its absolute temperature:
\[
L = 4\pi R^{2}\sigma_{SB} T^{4},
\]
where $L$ is luminosity, $R$ is the stellar radius, $T$ is surface temperature, and
$\sigma_{SB}$ is the Stefan--Boltzmann constant. The symbol $\pi$ appears here as the mathematical constant, in
keeping with the standard statement of the law, and elsewhere in the paper as
the propensity score $\pi(\cdot)$. This law implies a causal
relationship: temperature determines luminosity, and the causal direction is physically grounded. An object's temperature
dictates the energy it emits per unit area, making the Stefan--Boltzmann law a suitable
application for the generative learner. The $T^{4}$ dependence implies that even small
temperature changes produce large stellar luminosity changes: doubling temperature increases
luminosity by a factor of sixteen at constant radius. In empirical applications, departures from ideal black-body behavior are expected. Stellar surfaces are not perfectly isothermal, atmospheric processes alter the emergent radiation through absorption and re-emission, and both radius and effective temperature may be measured with error.  This uncertainty motivates the use of a flexible generative learner: the physical law provides sharp qualitative and quantitative restrictions, while the estimator is allowed to recover the corresponding effect structure from the observed data without imposing the \(T^4\)
functional form. 
 
Causal identification of $\theta(x)$ rests on the assumptions of
Subsection~\ref{subs_identification}.
Stable unit treatment value assumption may be plausible in this setting.
Stars are spatially isolated sources,
so assigning star $s$ to the hot or
cool temperature state does not change the luminosity of any other star, and the
luminosity of $s$ depends only on its own assignment. Overlap requires $0<\pi(x)<1$ at covariate profiles for which
$\theta(x)$ is to be interpreted: a star with a given set of characteristics must be able to appear in either treatment state. In practice, a researcher may inspect the estimated propensity scores and 
trim extreme values, so that the reported effects refer to stars
for which both temperature regimes remain possible rather than to
profiles that lie almost surely on one side of the treatment state.

Conditional unconfoundedness requires that, given $X_s$, the hot/cool
classification is independent of potential luminosities.
The restriction is plausible if residual variation in effective
temperature, among stars of similar radius, type, and color, is
generated by photospheric and environmental processes, such as surface
inhomogeneity, atmospheric absorption and re-emission, interstellar
reddening, and leftover differences in interior nuclear evolution that
shift the temperature of the radiating layers, rather than by an
independent determinant of radiated power.
Together these conditions identify $\theta(x)$ from the observed data: the average change
in brightness that would follow were a star of that covariate profile to
occupy the hot rather than the cool regime.

\subsection{Data}

We use publicly available stellar data that records physical and
spectral characteristics for $S=240$ stars, balanced across six types:
brown dwarfs, red dwarfs, white dwarfs, main-sequence stars, supergiants,
and hypergiants (40 stars per category).  \footnote{The dataset is available at
\href{https://github.com/YBIFoundation/Dataset/blob/main/Stars.csv}{\texttt{YBIFoundation/Dataset}}
and is described at
\href{https://www.kaggle.com/datasets/deepu1109/star-dataset/data}{\texttt{star-dataset}}. For the purpose of this empirical illustration, we treat stellar radius as an independently measured characteristic that remains unchanged under the hypothetical temperature intervention.}
For each star the file reports surface temperature in kelvin, luminosity
in solar units $L/L_{\odot}$, radius in solar units $R/R_{\odot}$,
absolute magnitude, color, and spectral class. We partition the sample into two halves of equal size, stratifying on
treatment status so that the treated proportion is preserved in each. Models are
estimated on the first half, and treatment effects are predicted out of sample on the
second.  

Let $T_s$ denote the surface temperature of star $s$.
Temperature is right-skewed, so a mean cutoff would be pulled toward a
small number of very hot stars.
We therefore take the sample median, $t^{\star}=\mathrm{median}(T)$, and
define the binary treatment
\begin{equation}
  D_s=\mathbf{1}\{T_s>t^{\star}\},
\end{equation}
which is the form required by the generative learner. $D_s=1$ labels a hot star and $D_s=0$ a cool star.
The outcome is luminosity, $Y_s=L_s$.
We report two covariate specifications.  The first conditions on stellar
radius together with stellar type and stellar color, and the second on
radius alone, in level and logarithm.   \ref{stellar_characteristics} visualizes the Hertzsprung-Russell (HR) diagram and the densities of the surface temperature, radius, and luminosity. Figures~\ref{fig:star-temperature}
and~\ref{fig:star-size-luminosity} in
\ref{stellar_characteristics} display a positive association
between temperature and luminosity that is most visible for the
main sequence stars.
The Hertzsprung--Russell diagram also makes clear that this
association is not uniform across types: white dwarfs are hot yet
faint, while giants are luminous at a wide range of temperatures.
The marginal distribution of luminosity is bimodal, with a mass of
stars near $L\approx 0$ and a second mass at very high $L$,
consistent with the separation between compact objects and giants.

\subsection{Findings}

\begin{figure}[H]
\centering
\begin{subfigure}[t]{0.48\textwidth}
    \centering
    \includegraphics[width=\linewidth]{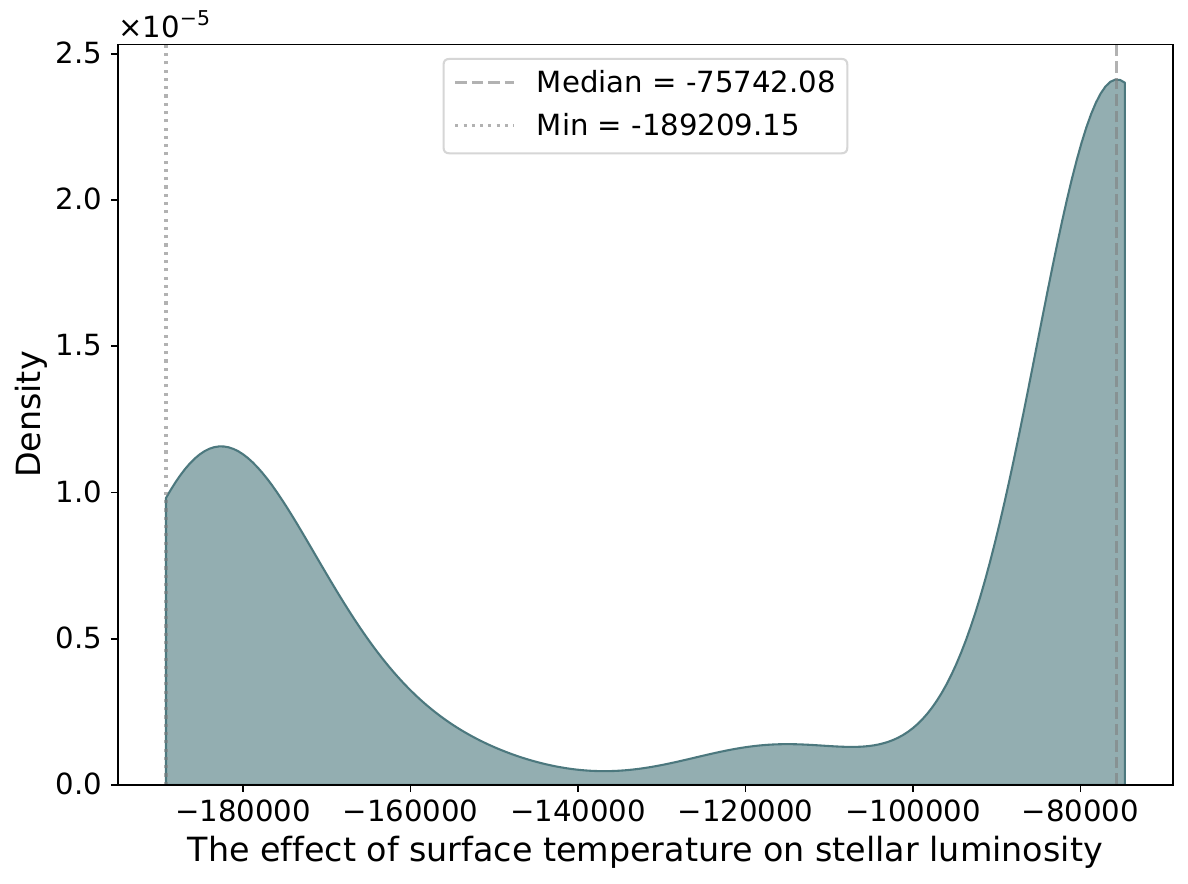}
    \caption{GRF (limited empirical overlap)}
    \label{fig:density_grf_full}
\end{subfigure}\hfill
\begin{subfigure}[t]{0.48\textwidth}
    \centering
    \includegraphics[width=\linewidth]{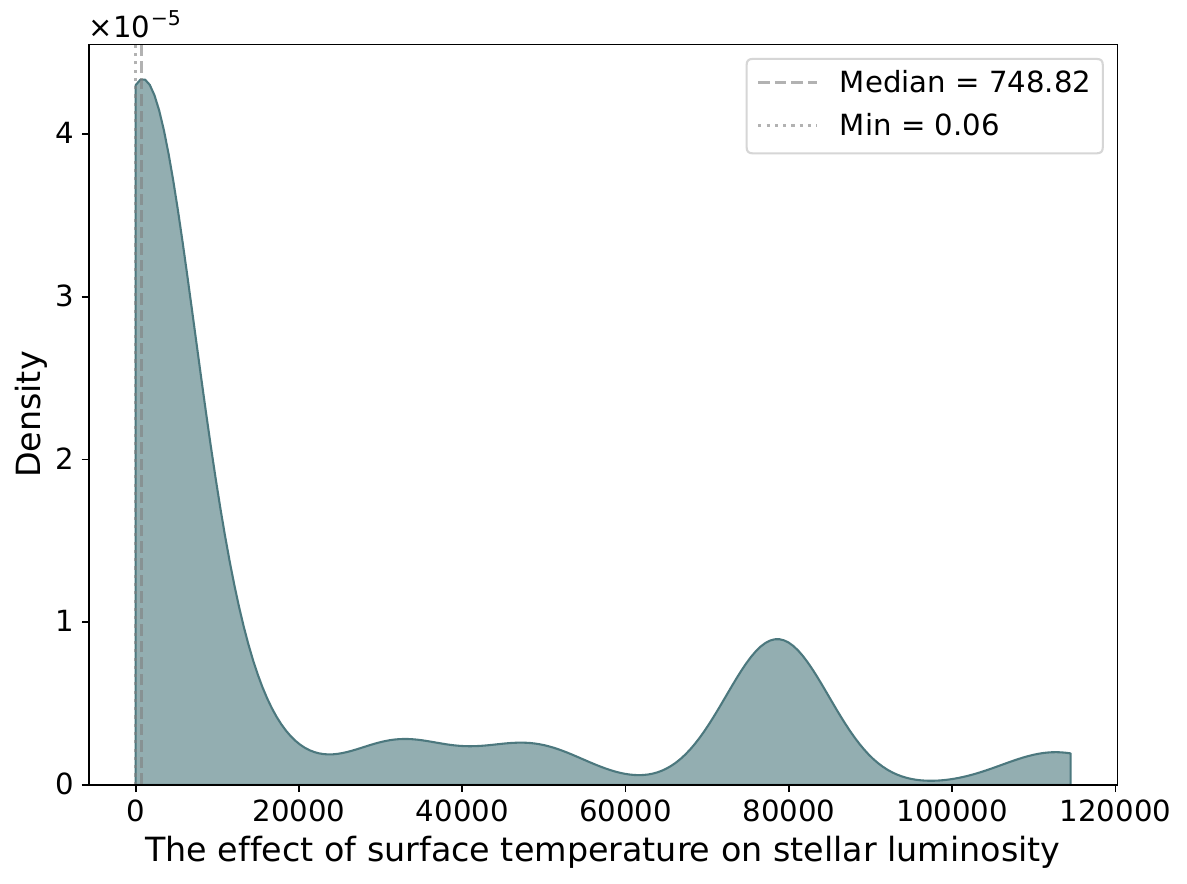}
    \caption{GenTE (limited empirical overlap)}
    \label{fig:density_gente_full}
\end{subfigure}

\vspace{0.8em}

\begin{subfigure}[t]{0.48\textwidth}
    \centering
    \includegraphics[width=\linewidth]{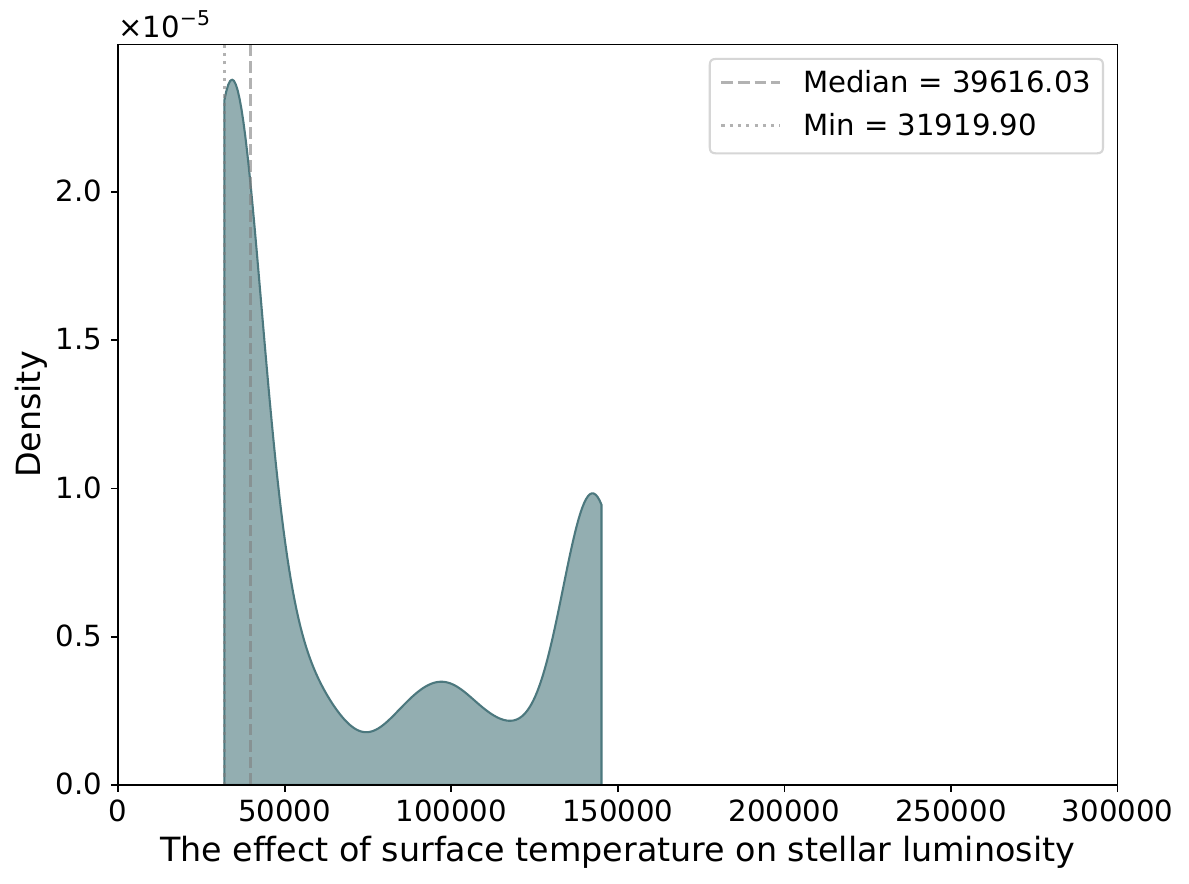}
    \caption{GRF (improved empirical overlap)}
    \label{fig:density_grf_ra}
\end{subfigure}\hfill
\begin{subfigure}[t]{0.48\textwidth}
    \centering
    \includegraphics[width=\linewidth]{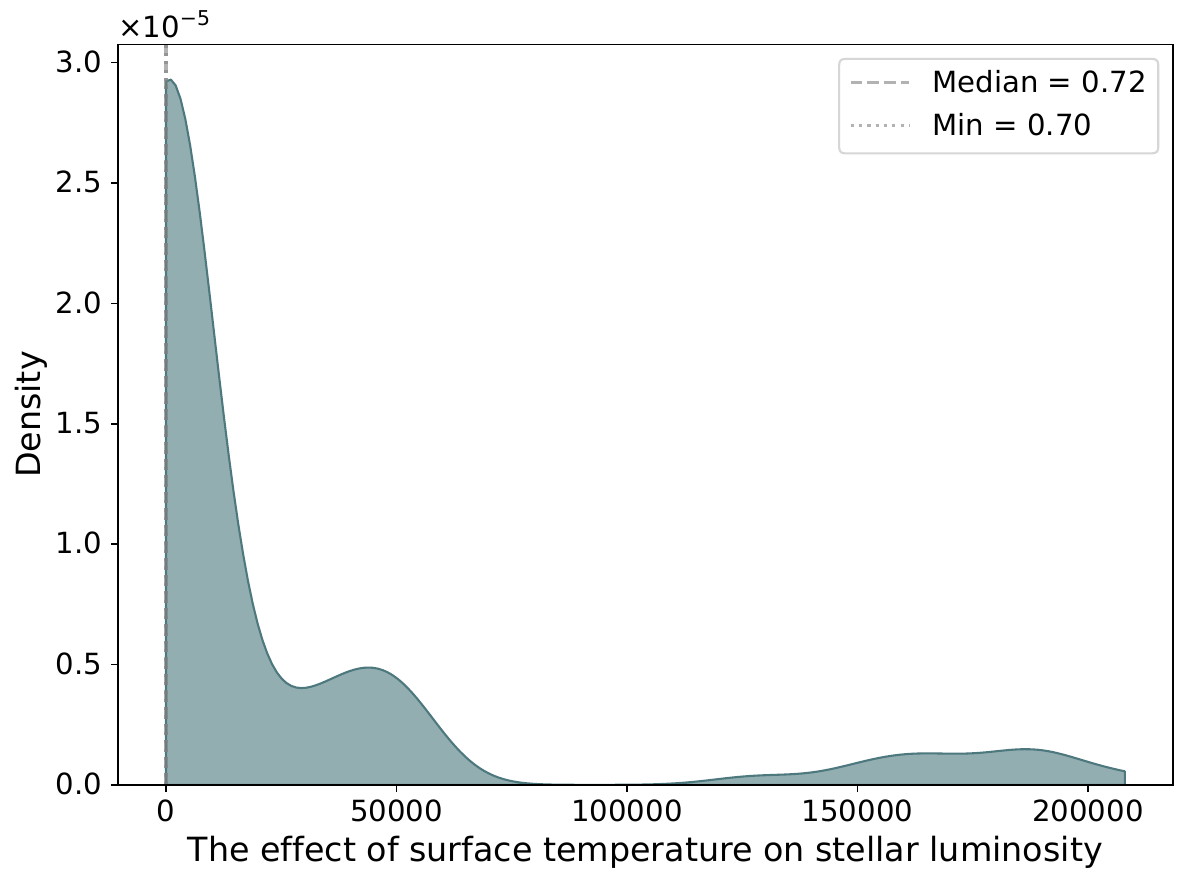}
    \caption{GenTE (improved empirical overlap)}
    \label{fig:density_gente_ra}
\end{subfigure}

\caption{Distributions of the estimated temperature effect on luminosity,
$\widehat{\theta}(X_s)$, on the held-out fold. The top row conditions on radius
together with stellar color and category ((a) and (b)). The bottom row
conditions on radius and the logarithm of radius alone ((c) and (d)). Each panel reports the median and the minimum of the estimated
effects.}
\label{fig:density_2x2}
\end{figure}

\autoref{fig:density_2x2} reports the estimated effects under two covariate
sets. The top row conditions on radius together with stellar color and
category. Stellar color is an alternative measurement of surface temperature: a star's
color is determined by its temperature through the blackbody spectrum. Treatment status is correspondingly close to a deterministic function of color.  Counting a
color as uninformative when the share of hot stars within it falls below $5\%$
or above $95\%$, $232$ of the $240$ stars ($97\%$ of the sample) lie in a
color that supplies no comparison between the two temperature regimes.
Conditioning on color therefore leaves almost no within-cell variation in treatment status, so the propensity score is close to zero or one at almost every covariate profile, resulting in extremely weak overlap. Stellar category compounds
this for a different reason: it separates stars by size rather than by
temperature, but three of its six classes lie entirely on one side of
$t^{\star}$, no brown or red dwarf in the catalog is hot and no white dwarf
is cool, so $120$ of the $240$ stars have no counterpart in the
opposite regime. Panel (a) of \autoref{fig:propensity_hist} in \ref{app_prop} visualizes this. On the other hand, radius varies within both temperature regimes and so does not
determine treatment status, and comparable stars remain available on
either side of $t^{\star}$ across the covariate support, as also verified by panel (b) of \autoref{fig:propensity_hist}. 

Degenerate propensity scores are consequential for estimators that divide by
them. Inverse propensity weights $D_s/\widehat{\pi}(X_s)$ and
$(1-D_s)/(1-\widehat{\pi}(X_s))$ diverge as $\widehat{\pi}(x)$ approaches the
boundary, and the residual-on-residual construction underlying the generalized
random forest \citep{wager2018estimation, athey2019generalized} carries the
treatment residual $D_s - \widehat{\pi}(X_s)$ in both the numerator and the
denominator of its local estimating equation. When that residual is numerically
zero for almost every observation, the estimate becomes a ratio of two
near-zero quantities determined by the few stars that depart from the pattern.
Panel~(a) of \autoref{fig:density_2x2} shows the consequence: the forest's estimates lie entirely below
zero, with a median of $-7.6\times10^{4}$, a sign the Stefan--Boltzmann law
excludes at every covariate value. GenTE parameterizes the treatment effect as
a component of the fitted conditional quantile function rather than as a ratio
of residuals, and its estimates remain positive under the same specification
(panel~(b), median $749$, minimum $0.06$). This difference should be read as evidence on sensitivity within a particular
design rather than as a general robustness property. Where a covariate profile
admits only one treatment state, the corresponding potential outcome is
unobserved at any sample size and $\theta(x)$ is not identified there for either
estimator. The values each method reports at such profiles reflect its
extrapolation behavior rather than information in the data. What the comparison
establishes is narrower, and concerns the direction that extrapolation takes:
GenTE's estimates remain within the region the Stefan--Boltzmann law admits,
whereas the estimates of the forest do not. The bottom row (panels (c) and (d)) conditions on radius (and the logarithm of radius) alone, the one covariate that varies within
temperature regimes, so that $\widehat{\pi}(x)$ stays away from zero and one
across the covariate support. Both methods produce positive estimated effects under the reduced covariate specification. This suggests that the extreme negative estimates obtained under the richer specification are associated with the choice of adjustment variables and the resulting lack of empirical overlap, although the comparison does not isolate the precise source of the discrepancy.

What remains is the range of
heterogeneity each can express. The Stefan--Boltzmann law writes luminosity as $L_s=4\pi\sigma_{SB} R_s^{2}T_s^{4}$. Under the law the potential luminosities in the hot
and cool temperature states are given as
\begin{align*}
L_s(1)
&=
4\pi\sigma_{SB} R_s^{2}\,T_s(1)^{4},
&
L_s(0)
&=
4\pi\sigma_{SB} R_s^{2}\,T_s(0)^{4}.
\end{align*}
Assuming that stellar radius is a pre-treatment characteristic unaffected by the temperature, $L_s(1)$ and
$L_s(0)$ differ only through $T_s(1)^{4}$ and $T_s(0)^{4}$.
The CATE is therefore
\begin{align}\label{theta_sb}
\theta(x)
=
\E\bigl[L_s(1)-L_s(0)\mid X_s=x\bigr]
=
4\pi\sigma_{SB} R(x)^{2}
\Bigl\{\E\bigl[T_s(1)^{4}\mid X_s = x\bigr]
-
\E\bigl[T_s(0)^{4}\mid X_s = x\bigr]\Bigr\}.
\end{align}

Given the substantial variation in stellar radius in the catalog, the \(R^2\) scaling implies the potential for treatment effects to vary over several orders of magnitude, provided that differences in the temperature contrast do not offset the radius effect. GenTE's estimates span $5.5$ decades, from
$0.70$ to $2.1\times10^{5}$, and retain the bimodal shape of the underlying
distribution of luminosity, visible in Panel (d) of \autoref{fig:density_2x2}. The estimates of GRF span $0.66$ decades, from
$3.2\times10^{4}$ to $1.5\times10^{5}$, seen through Panel (c) of \autoref{fig:density_2x2}. Forest predictions are locally weighted
averages of neighboring pseudo-outcomes, so with $S = 120$ training stars and an
outcome dominated by a small number of giants, the attainable range is limited by
the leaf averages that can be formed.

\begin{figure}[H]
    \centering
    \begin{subfigure}{0.32\textwidth}
        \centering
        \includegraphics[width=\textwidth]{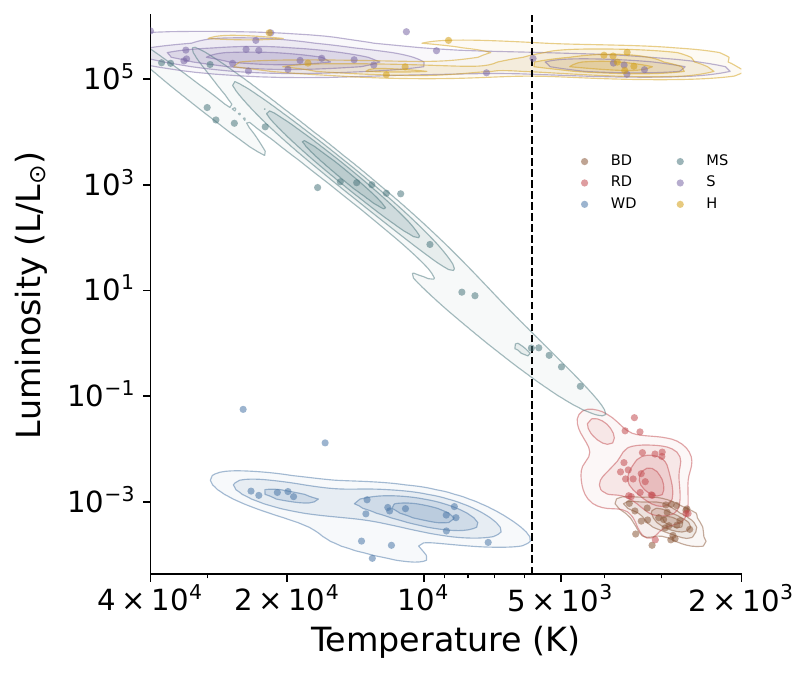}
        \caption{Luminosity}
        \label{fig:stars_heatmap_TL}
    \end{subfigure}
    \hfill
    \begin{subfigure}{0.32\textwidth}
        \centering
        \includegraphics[width=\textwidth]{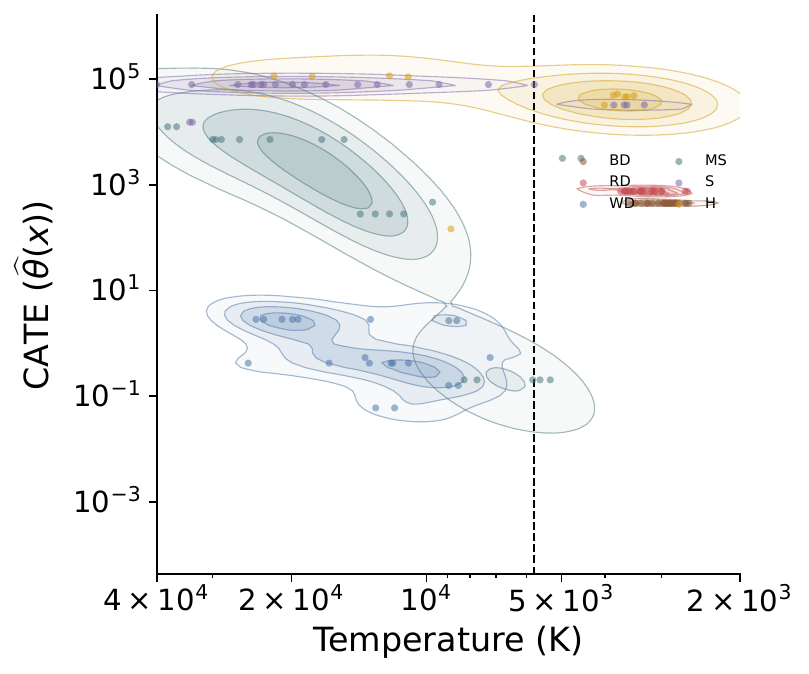}
        \caption{$\widehat{\theta}(x)$ (limited empirical overlap)}
        \label{fig:stars_heatmap_Ttau}
    \end{subfigure}
    \hfill
    \begin{subfigure}{0.32\textwidth}
        \centering
        \includegraphics[width=\textwidth]{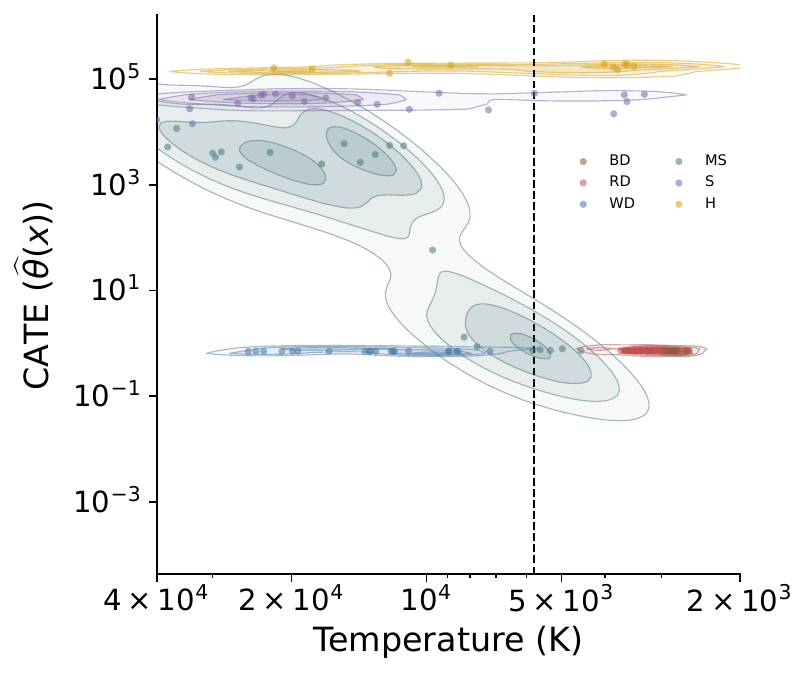}
        \caption{$\widehat{\theta}(x)$ (improved empirical overlap)}
        \label{fig:stars_heatmap_Ttau_ra}
    \end{subfigure}
    \caption{Hertzsprung--Russell diagram on the test fold. Panel~(a) plots
    observed luminosity against temperature. Panels~(b) and~(c) plot the
    estimated temperature effect on luminosity, $\widehat{\theta}(x)$, against
    temperature, conditioning respectively on radius together with stellar
    color and category in (b), and on radius and the logarithm of radius in (c). Each
    stellar class is shown as a soft density map with points overlaid. Classes
    are brown dwarf (BD), red dwarf (RD), white dwarf (WD), main sequence (MS),
    supergiant (S), and hypergiant (H). The temperature axis increases from
    right to left in each panel.}
    \label{fig_stars_hr}
\end{figure}

We also note that the term in braces of \eqref{theta_sb} is a temperature contrast and the surface area scales it. The same shift in $T^{4}$ therefore produces a small change in $L$ for a compact star and a large one for an extended star. Specifically, a star ten times larger gets a hundred times the effect from the identical temperature contrast.  The effect distribution inherits the split between faint dwarfs and
luminous giants, as shown by \autoref{fig_stars_hr}. The figure displays the same stars three times: observed luminosity
against temperature in panel~(a), and the estimated effect of a temperature shift from the
cool to the hot regime in panels~(b) and~(c). The estimated effects order by size: the compact classes, brown, red and white dwarfs, have the smallest effects, main-sequence stars follow, and
supergiants and hypergiants have the largest. The ordering across the three
dwarf classes in (b) should be read with the caveat of the preceding paragraph: none of them contains both temperature regimes, so their effects are extrapolated rather
than estimated.

Testing a physical law against observational estimates is difficult in practice.
The law is a statement about a continuous relationship,
$L = 4\pi\sigma_{SB} R^{2}T^{4}$, whereas the estimand here is a contrast between two
temperature regimes, and the sample is small and does not itself reproduce the
law exactly. We therefore ask the weaker question of whether the estimated
effects satisfy the restrictions the law implies. The law yields two direct restrictions and a further empirical implication when combined with the radius distribution in the catalog. First,
luminosity is strictly increasing in temperature at fixed radius, so
$\theta(x) > 0$ at every covariate value. GenTE's smallest estimate on the
held-out fold is $0.70$ solar luminosities, and no star receives a negative
effect (panel (d) of \autoref{fig:density_2x2}). Second, holding the temperature contrast fixed, the treatment effect increases with stellar radius through the emitting-area, so effects are expected to be ordered by stellar size. The estimated effects respect this pattern. The
compact classes have the smallest estimated effects, main-sequence stars follow,
and supergiants and hypergiants have the largest (\autoref{fig_stars_hr}). Third,  the estimated effect
distribution inherits the separation of the catalog by radius.  Its
two modes correspond to the compact and the extended classes, and they
align with the faint-dwarf and luminous-giant masses of the empirical
luminosity density (\ref{stellar_characteristics}).  \footnote{\autoref{fig:lfit_scatter} of \ref{app_pred_perf} additionally reports the predicted stellar luminosity on the
held-out stars: predicted luminosity tracks the observed values across
the luminous range under both covariate sets and flattens where the dwarf classes have no counterfactual
support.}  These findings are supporting rather than
confirmatory. They establish that the estimated effect surface lies where the
law admits and varies with radius as the law prescribes, not that the
$R^{2}T^{4}$ form has been recovered.

\section{Discussion and Conclusion}

We propose GenTE, a generative learner that estimates
conditional average treatment effects by learning the conditional quantile
functions of the potential outcomes rather than their conditional means.
The estimator is a multi-head neural network that fits the propensity score,
the baseline outcome and the treatment effect under a single objective,
combining covariate features with a cosine quantile embedding by
element-wise multiplication, in line with the implicit quantile network of \cite{dabney2018implicit}.  Two exercises assess the estimator.  In the Monte Carlo designs, where the oracle CATE is known, GenTE attains the
lowest out-of-sample mean squared error in 30 of 36 cases, with the
widest margins where the effect is a nonlinear function of the
covariates.  We apply the learner to the Stefan--Boltzmann law, which states that the
power radiated by a black body rises with the fourth power of its
absolute temperature, so that a hotter star of a given size is
necessarily the brighter one.  The law implies the restrictions a credible estimate of the treatment effect should meet: the effect
must be positive at every covariate profile, it is increasing in the
emitting area, and it therefore is expected to vary over several
orders of magnitude across a catalog whose radii span several
decades.  The estimated conditional average treatment effects satisfy
all three implications.  A law of this kind supplies an external
standard against which estimates can be judged.

Three features of the architecture plausibly contribute to these gains.
The covariate network $\psi(x)$ learns a flexible representation of the covariate vector that may exploit low-dimensional structure in the target function, potentially reducing the effective complexity of the estimation problem \citep{jiang2017learning} that tree-based and kernel
methods must solve over the full covariate space
\citep{coppejans2000breaking}.  The quantile formulation exploits the
fact that a conditional quantile function is itself a composition
\citep{parzen2004quantile, polson2025generative}, and deep ReLU networks
attain rates on compositional classes that depend on the intrinsic
dimensions of each oracle compositional layer rather than on the ambient covariate dimension
(Propositions~\ref{prop:sh} and~\ref{prop:padilla}).  And the
multiplicative form $g_{d}(\psi(x)\odot\varphi(q))$ supplies
the interaction between covariates and quantile level by construction,
where an additive or concatenated alternative would have to learn it
\citep{schmidt-hieber2020nonparametric, dabney2018implicit}.

The proposed construction has a probabilistic interpretation.  Evaluating the
generator at a uniform draw is the inverse-transform sampler: a draw
from a fixed reference distribution is carried into a draw from the
target conditional law.  Conditioning is then transport rather than
integration, since the covariates and the treatment state deform the
reference distribution into $F^{-1}_{Y(d)\mid X=x}$, and a new
conditioning argument updates that map rather than requiring a fresh
calculation \citep{polson2024generativea, polson2025generative}.  The
prior-to-posterior step of Bayesian inference has the same shape.  What
changes is the distribution, and the information enters through the map
that generates draws from it.  The difference is where the map comes
from: it is learned from data rather than derived from a likelihood,
which is why a single fit delivers the whole conditional law of each
potential outcome, and with it any functional of that law.

Several limitations point to extensions.  Identification rests on
selection on observable covariates.  Extending the
architecture to accommodate endogeneity, through an instrumental variable approach or a control
function approach, would widen its scope.  Asymptotic theory for the estimator
remains open: rates for the fitted quantile surface and valid coverage
for $\widehat\theta(x)$ would enable researchers and policymakers to draw inference on the parameter of interest.  The proposed algorithm is currently formulated for binary treatments. Extending the framework to multi-valued or continuous treatments constitutes a natural direction for future research and would require a corresponding reformulation of the estimand and training objective to accommodate marginal quantile treatment effects. Finally, the model provides a convenient framework for testing conditional independence between a binary treatment and an outcome by assessing whether the quantile-specific treatment effect is uniformly zero across the outcome distribution, without requiring an explicit partition of either the covariate space or the outcome space. Developing a randomization inference  for such a test will be a useful direction for future research.

\bibliographystyle{apalike}
\bibliography{Econ}

\begin{appendix}

\section{Loss Function in a Deep Learner}
\label{app:loss}

Training GenTE in \autoref{fig:architecture} requires a single objective that
serves three purposes at once: it must fit the treatment assignment mechanism,
pin down the level of the outcome, and recover the shape of the conditional
outcome distribution. The objective therefore combines a treatment-assignment
loss with three outcome-fitting loss functions,
\begin{equation}
\mathcal L(w;\, q, q')
=
\mathcal L_{\mathrm{BCE}}(w)
+
\lambda_0\,\mathcal L_{\mathrm{MAE}}(w;q)
+
\lambda_1\,\mathcal L_{\mathrm{mono}}(w;q,q')
+
\lambda_2\,\mathcal L_{\mathrm{pinball}}(w;q),
\label{eq:total-loss}
\end{equation}
where $w$ collects all network parameters and $\lambda_0,\lambda_1,\lambda_2\ge 0$
are fixed weights. Quantile levels are not tuning parameters but arguments of
the loss: two levels $q, q' \sim \text{Unif}(0,1)$ are drawn independently at
each iteration, the network is evaluated at both, and the gradient is computed
on the full sample, so the optimization targets
\begin{equation}
\hat w \in \arg\min_{w}\;
\mathbb{E}_{q, q'}\bigl[\mathcal L(w;\,q, q')\bigr].
\label{eq:objective}
\end{equation}
The first and third outcome terms use the evaluation at $q$ alone, the second
requires both, since a quantile crossing can only be detected by comparing two levels.
Because updates use the full sample, every iteration is a single gradient step
in which all observations share the drawn levels, and the expectation in
\eqref{eq:objective} is approximated across iterations rather than within them.
One network is therefore trained across a continuum of quantile levels rather
than a fixed grid, and no separate model has to be estimated for each level of
interest.

\section*{Treatment assignment}

The propensity subnetwork maps covariates to an embedding, and a small output
network $\mathcal{G}_{\pi}$ maps that embedding to a scalar index and then to a
treatment probability,
\begin{align}
e_\pi(X_s) &\in \mathbb{R}^{d_\pi} && \text{(propensity embedding)},\\
\eta_s &:= \mathcal{G}_{\pi}\bigl(e_\pi(X_s)\bigr) && \text{(index)},\\
\hat\pi(X_s) &:= \Lambda(\eta_s) \;\approx\; \Pr\bigl(D_s = 1 \mid X_s\bigr)
&& \text{(propensity score)} ,
\end{align}
with $\Lambda$ the logistic function. The treatment assignment model is fitted by
minimizing the negative log-likelihood of the treatment indicator under this logistic link,
equivalently the binary cross-entropy
\begin{equation}
\mathcal L_{\mathrm{BCE}}
=
-\frac{1}{S}\sum_{s=1}^{S}
\Bigl[
D_s \log \hat\pi(X_s) + (1-D_s)\log\bigl(1-\hat\pi(X_s)\bigr)
\Bigr].
\label{eq:bce}
\end{equation}
This term enters with unit weight. Its role is not only to estimate
$\pi(\cdot)$: the embedding $e_\pi(X_s)$ is passed to the baseline as its second
argument, so \eqref{eq:bce} is what makes that input informative about the
treatment assignment mechanism and thereby implements the reparameterization in
\eqref{eq:model}.

\section*{Outcome modeling}

The output head returns two coordinates, which play distinct roles. Writing
$\hat\mu_s := \hat\mu\bigl(X_s, e_\pi(X_s), q\bigr)$ in the specification
reported here (and $\hat\mu_s := \hat\mu\bigl(X_s, e_\pi(X_s)\bigr)$, without
the quantile argument, when the target parameter is the conditional median
effect), they are
\begin{equation}
\hat Y^{\mathrm{a}}_s(q) := \hat\mu_s + \hat\theta_{\mathrm a}(X_s,q)\,D_s,
\qquad
\hat Y^{\mathrm{q}}_s(q) := \hat\mu_s + \hat\theta(X_s,q)\,D_s ,
\label{eq:two-heads}
\end{equation}
where the second coordinate carries the quantile-indexed effect
$\hat\theta(X_s,q)$ of \eqref{eq:model-components} and the first is a prediction of the
outcome level whose target does not vary with $q$. Denote the corresponding
residuals by $e^{\mathrm a}_s(q) := Y_s - \hat Y^{\mathrm a}_s(q)$ and
$e^{\mathrm q}_s(q) := Y_s - \hat Y^{\mathrm q}_s(q)$, so that negative residuals
indicate overprediction and positive residuals underprediction.

The first coordinate is fitted by the mean absolute deviation,
\begin{equation}
\mathcal L_{\mathrm{MAE}}(q)
=
\frac{1}{S}\sum_{s=1}^{S}\bigl|e^{\mathrm a}_s(q)\bigr| .
\label{eq:mae}
\end{equation}
The
baseline $\hat\mu$ and the covariate encoders are shared by both output heads, so
this term supplies them with a target that is invariant to the drawn quantile level and
encourages this coordinate to represent a \(q\)-invariant location target, in the spirit of least absolute
deviations. Squared error would serve the same purpose but would allow a small
number of extreme outcomes to dominate the fit. In this paper, we fix \(\lambda_0 = 0\). Setting $\lambda_0 > 0$
is appropriate when the goal is to estimate conditional median effects rather than averages.

The second term discourages crossings of the fitted conditional quantile curves. For the ordered pair of drawn levels, taken without loss of generality as \(q<q'\),

\begin{align}\label{eq:mono}
 \mathcal L_{\mathrm{mono}}(q,q')
=
\frac{1}{S}\sum_{s=1}^{S}
\max\!\Bigl\{0,\,
\hat Y^{\mathrm q}_s(q)-\hat Y^{\mathrm q}_s(q')
\Bigr\},
\end{align}

\noindent which is zero whenever the fitted quantile function for the treatment state observed for unit \(s\) is non-decreasing between the two levels, and positive whenever the fitted value at the lower quantile exceeds that at the higher quantile. Thus, control observations penalize crossings of the fitted control quantile function, whereas treated observations penalize crossings of the fitted treated quantile function. Because conditional quantile functions are non-decreasing by construction, this term acts as a soft regularization toward the corresponding monotonicity restriction rather than imposing it directly.

The third term is the quantile loss function (check function),
\begin{equation}
\mathcal L_{\mathrm{pinball}}(q)
=
\frac{1}{S}\sum_{s=1}^{S}
\rho_q\bigl(e^{\mathrm q}_s(q)\bigr),
\qquad
\rho_q(u) = \max\bigl\{q\,u,\ (q-1)\,u\bigr\} = u\Bigl(q - \mathbf 1\{u<0\}\Bigr) ,
\label{eq:pinball}
\end{equation}
which weights underprediction by $q$ and overprediction by $1-q$ and is
minimized when $\hat Y^{\mathrm q}_s(q)$ equals the conditional $q$-quantile of
$Y_s$ given $(X_s, D_s)$. This is the term that
identifies the quantile structure, and it carries the largest weight among the outcome terms.

The three outcome terms are complementary rather than competing. The
penalty \eqref{eq:mono} is inactive at any non-crossing configuration, so it
does not shift the target of \eqref{eq:pinball}. For the remaining two, since
\begin{equation}
\int_0^1 \rho_q(u)\,dq = u \int_0^1 \Bigl(q - \mathbf 1\{u<0\}\Bigr) dq
= u\Bigl(\tfrac12 - \mathbf 1\{u<0\}\Bigr)
= \tfrac12\,|u| ,
\label{eq:pinball-avg}
\end{equation}
averaging the check function over $q \sim \text{Unif}(0,1)$ returns one half of
the absolute deviation. Under the objective \eqref{eq:objective} the two
criteria therefore point to the same target for any component that does not
depend on $q$. This is relevant only in the median specification, in which the
baseline carries no quantile argument: there both coordinates in
\eqref{eq:two-heads} reduce to $\hat\mu_s$ on control units, so those
observations drive the baseline to the conditional median of the control
outcome, while on treated units the effect coordinate absorbs the remainder.
In the specification reported here $\lambda_0 = 0$ and the baseline is
quantile-indexed, so \eqref{eq:mae} is inactive and the control units instead
drive $\hat\mu\bigl(X_s, e_\pi(X_s), q\bigr)$ to the conditional $q$-quantile
of the control outcome under \eqref{eq:pinball}.

\section*{Implementation}

Because the quantile levels are redrawn at each iteration, the fitted
$\widehat{\theta}(X_s,q)$ is available at every level in $(0,1)$ after a single
training run. We set $(\lambda_0,\lambda_1,\lambda_2) = (0,\,0.1,\,0.6)$ in this
paper. Among the outcome-loss components, the largest weight is assigned to the check function. Updates in the network use the full sample at each iteration, the weights of the subnetworks are
optimized by adaptive moment estimation. In each experiment, $3$ models trained from
different random initializations are averaged to reduce the sensitivity of the
estimates to any single run.

\section{Scatterplots of CATEs}\label{app_scatters}

\begin{figure}[H]
\centering

\begin{subfigure}[t]{0.48\textwidth}
  \centering  \includegraphics[width=\linewidth]{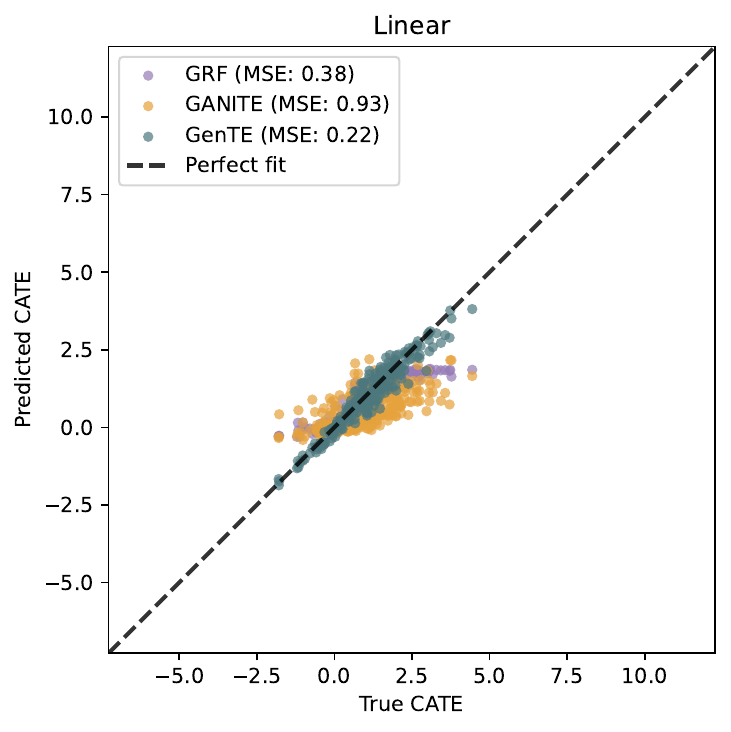}
  \caption{Linear}
\end{subfigure}\hfill
\begin{subfigure}[t]{0.48\textwidth}
  \centering
  \includegraphics[width=\linewidth]{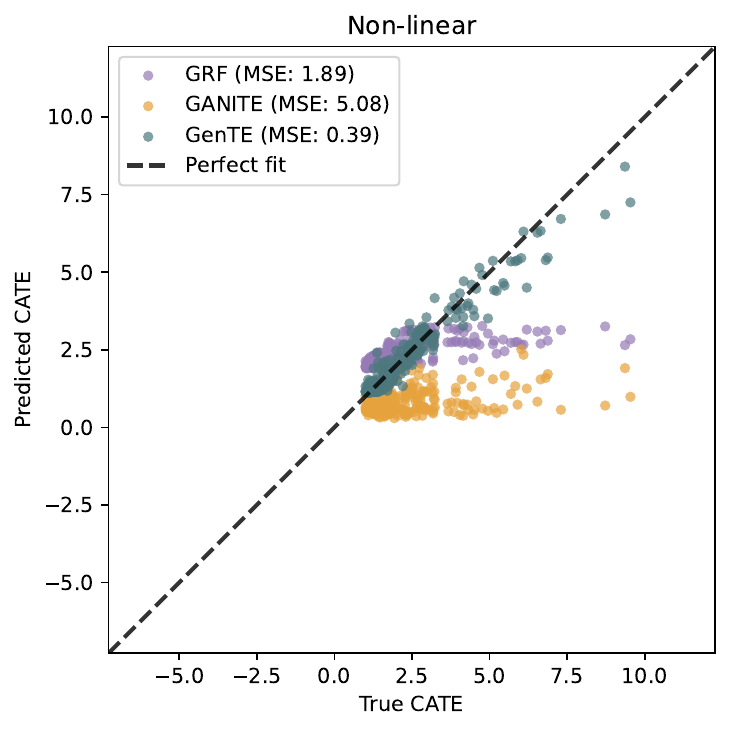}
  \caption{Nonlinear}
\end{subfigure}

\vspace{0.6em}

\begin{subfigure}[t]{0.48\textwidth}
  \centering
  \includegraphics[width=\linewidth]{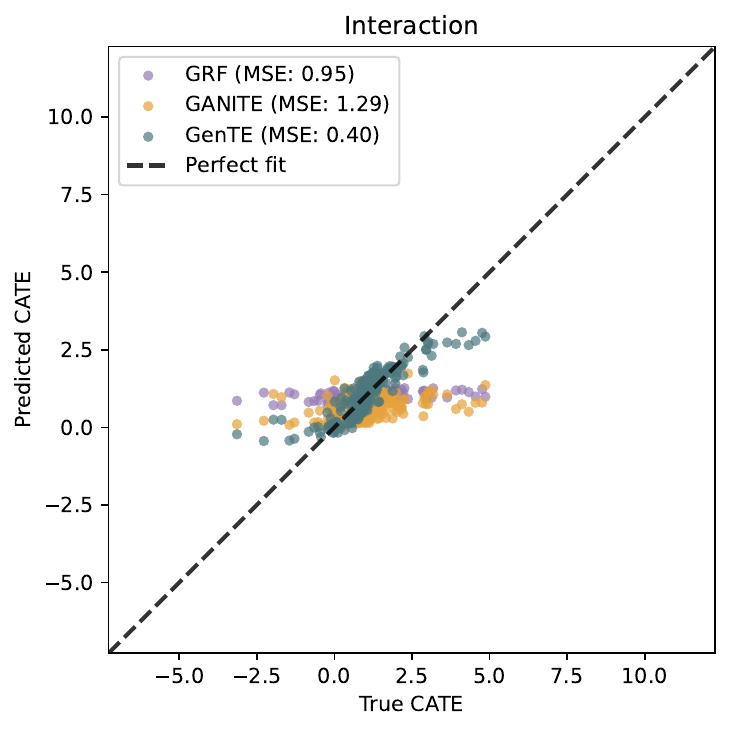}
  \caption{Interaction}
\end{subfigure}\hfill
\begin{subfigure}[t]{0.48\textwidth}
  \centering
\includegraphics[width=\linewidth]{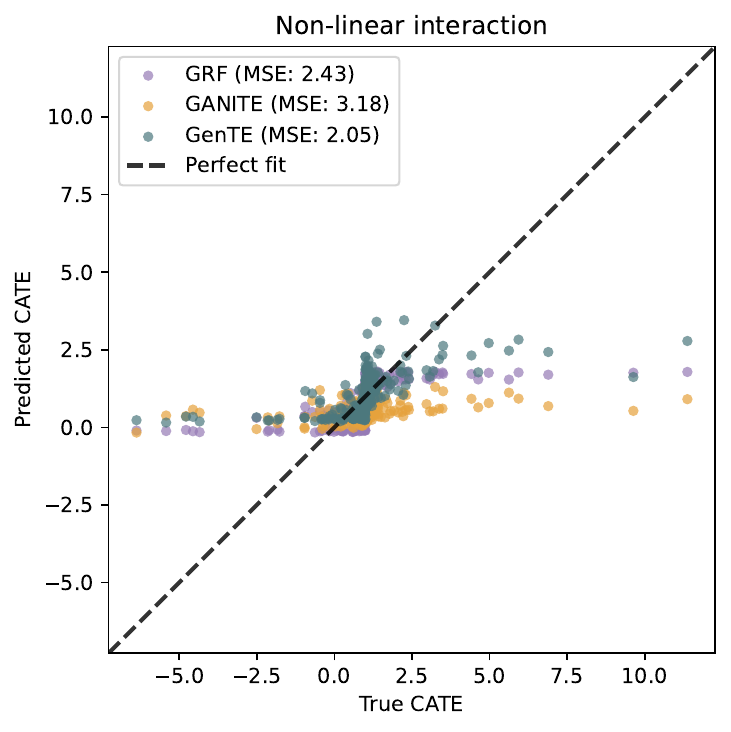}
  \caption{Nonlinear interaction}
\end{subfigure}

\caption{Scatterplots of the conditional average treatment effect (CATE) across estimators, evaluated on the test set and averaged over 100 Monte Carlo replications. GenTE denotes the generative treatment effect learner proposed in this article. GRF denotes the generalized random forest estimator \citep{athey2019generalized}. GANITE represents generative adversarial networks for individualized treatment effects \citep{yoon2018ganite}. True CATE is the simulated oracle conditional average treatment effect. Each simulated experiment uses a sample size of 500. }
\label{fig:sim_comparison_ganite_scatter}
\end{figure}

\section{A T-Learner Benchmark}\label{app_T_learner}

\begin{table}[H]
\centering
\caption{Mean squared error of the GenTE estimate of the conditional average
treatment effect, by effect type, sample size $S$, and covariate dimension $p$, averaged over 100 Monte Carlo experiments.
GenTE (separate) fits the treated and
control conditional quantile functions in separate networks and takes their
difference. GenTE (joint, grid) is the specification reported in the main text: a
single fit of the baseline, effect and propensity subnetworks under the
composite loss of \ref{app:loss}, with the CATE obtained by averaging
$\hat\theta(x,q)$ over a grid of 19 quantile levels. GenTE (joint, MC)
is that same fit, summarized instead by averaging $\hat\theta(x,q)$ over 500
quantile levels drawn independently from $\mathrm{Unif}(0,1)$ for each ensemble
member. The two joint columns therefore differ only in how the integral in
\eqref{eq:cate-identified} is approximated, and not in the estimator. Each cell
is the mean MSE against the oracle CATE with the Monte Carlo standard deviation
in parentheses, over 100 replications. \bt{Dark blue} marks the cells
in which the separate-arm estimator attains a lower MSE than joint estimation. }
\label{tab:gente_sep_joint}
\resizebox{0.8\textwidth}{!}{%
\begin{tabular}{lrrrrrrrrr}
\toprule
& \multicolumn{3}{c}{GenTE (separate)} & \multicolumn{3}{c}{GenTE (joint, grid)} & \multicolumn{3}{c}{GenTE (joint, MC)} \\
\cmidrule(lr){2-4}\cmidrule(lr){5-7}\cmidrule(lr){8-10}
Sample size & $p{=}2$ & $p{=}4$ & $p{=}8$ & $p{=}2$ & $p{=}4$ & $p{=}8$ & $p{=}2$ & $p{=}4$ & $p{=}8$ \\
\midrule
\multicolumn{10}{l}{Linear} \\
\midrule
200 & 0.722 & 1.124 & 1.151 & 0.107 & 0.345 & 0.712 & 0.106 & 0.346 & 0.715 \\
  & (0.163) & (0.335) & (0.330) & (0.067) & (0.165) & (0.380) & (0.067) & (0.165) & (0.381) \\
500 & 0.676 & 1.068 & 1.124 & 0.064 & 0.222 & 0.409 & 0.063 & 0.223 & 0.410 \\
  & (0.122) & (0.200) & (0.145) & (0.046) & (0.138) & (0.232) & (0.046) & (0.138) & (0.233) \\
1000 & 0.680 & 1.054 & 1.115 & 0.053 & 0.170 & 0.302 & 0.053 & 0.171 & 0.303 \\
  & (0.111) & (0.185) & (0.142) & (0.046) & (0.101) & (0.147) & (0.047) & (0.102) & (0.148) \\
\addlinespace
\midrule
\multicolumn{10}{l}{Non-linear} \\
\midrule
200 & 2.435 & 2.494 & 2.647 & 0.346 & 0.660 & 1.457 & 0.347 & 0.658 & 1.461 \\
  & (0.923) & (0.731) & (0.923) & (0.297) & (0.393) & (0.633) & (0.297) & (0.392) & (0.635) \\
500 & 2.321 & 2.376 & 2.639 & 0.174 & 0.348 & 0.688 & 0.174 & 0.345 & 0.687 \\
  & (0.504) & (0.543) & (0.562) & (0.100) & (0.231) & (0.305) & (0.100) & (0.231) & (0.306) \\
1000 & 2.304 & 2.418 & 2.629 & 0.137 & 0.275 & 0.565 & 0.136 & 0.272 & 0.563 \\
  & (0.375) & (0.373) & (0.370) & (0.063) & (0.119) & (0.259) & (0.063) & (0.119) & (0.261) \\
\addlinespace
\midrule
\multicolumn{10}{l}{Interaction} \\
\midrule
200 & 0.991 & 1.074 & \bt{1.069} & 0.358 & 0.578 & 1.169 & 0.357 & 0.578 & 1.172 \\
  & (0.302) & (0.334) & (0.373) & (0.206) & (0.258) & (0.453) & (0.206) & (0.259) & (0.454) \\
500 & 0.949 & 1.033 & 1.032 & 0.227 & 0.377 & 0.666 & 0.227 & 0.376 & 0.668 \\
  & (0.175) & (0.206) & (0.174) & (0.108) & (0.142) & (0.276) & (0.108) & (0.142) & (0.279) \\
1000 & 0.962 & 1.025 & 1.046 & 0.203 & 0.308 & 0.521 & 0.202 & 0.308 & 0.522 \\
  & (0.146) & (0.144) & (0.145) & (0.094) & (0.115) & (0.191) & (0.094) & (0.116) & (0.192) \\
\addlinespace
\midrule
\multicolumn{10}{l}{Non-linear interaction} \\
\midrule
200 & 2.893 & 2.645 & \bt{2.792} & 1.954 & 2.155 & 3.197 & 1.953 & 2.154 & 3.204 \\
  & (1.906) & (1.402) & (1.468) & (1.640) & (1.139) & (1.550) & (1.636) & (1.138) & (1.553) \\
500 & 2.647 & 2.559 & 2.906 & 1.403 & 1.632 & 2.519 & 1.400 & 1.632 & 2.523 \\
  & (1.015) & (0.865) & (0.922) & (0.930) & (0.787) & (1.019) & (0.928) & (0.786) & (1.022) \\
1000 & 2.647 & 2.629 & 2.940 & 1.364 & 1.525 & 2.231 & 1.360 & 1.524 & 2.231 \\
  & (0.736) & (0.673) & (0.607) & (0.641) & (0.641) & (0.602) & (0.641) & (0.641) & (0.601) \\
\bottomrule
\end{tabular}%
}
\end{table}

\newpage
\section{Robustness to Treatment Assignment Mechanisms and Signal-to-Noise Ratios}\label{app_assignment}

The Monte Carlo designs in the main text vary the complexity of
\(\theta(x)\) while holding the propensity score fixed at a linear
logistic rule. Here we reverse that comparison. The outcome model and
the CATE are held fixed at the most demanding specification,
\begin{equation}
  Y_s=\mu(X_s)+\theta(X_s)\,D_s+\varepsilon_s,
  \qquad
  \theta(x)=1+x_1^{2}x_2,
\end{equation}
with \(X_s\sim\mathcal N(0,I_p)\),
\(\mu(x)=x^{\top}\beta_{\mu}\),
\(\beta_{\mu,j}\stackrel{\mathrm{iid}}{\sim}\mathcal N(0,1/4)\) and
\(\varepsilon_s \sim \mathcal N(0,1/4)\). We fix the number of covariates \(p=20\), so that most
coordinates of \(X_s\) enter only through the baseline
\(\mu\), and the sample size \(S = 500\). Treatment is Bernoulli,
\(D_s\mid X_s=x\sim\mathrm{Bernoulli}\bigl(\pi(x)\bigr)\), and we vary only
the assignment mechanism \(\pi(x) = \mathbb P (D = 1 | X = x)\).

\begin{enumerate}
\item Randomized controlled trial (RCT): assignment is independent of the covariates,
  \(\pi(x)=1/2\). This is the unconfounded benchmark.

\item Linear logit: the propensity is the logistic of a linear
  index, as in the main observational design,
  \begin{equation}
    \pi(x)=\Lambda\bigl(0.3\,x_1-0.2\,x_2+0.1\,x_3\bigr),
    \qquad
    \Lambda(z)=(1+e^{-z})^{-1}.
  \end{equation}
  Confounding is therefore confined to the first three coordinates.

\item Nonlinear logit: the same logistic link is retained, but
  the index has the same quadratic-interaction shape as the CATE,
  \begin{equation}
    \pi(x)=\Lambda\bigl(0.3\,x_1^{2}x_2+0.1\,x_3\bigr),
  \end{equation}
  Selection into treatment now depends on the same nonlinear feature
  that determines \(\theta(x)\).

\item Trigonometric: the propensity is bounded and
  non-monotone,
  \begin{equation}
    \pi(x)=\tfrac12+\tfrac25\sin(x_1)\cos(x_2)\in[0.1,0.9].
  \end{equation}
  Unlike the linear-logit specification, this treatment assignment mechanism is non-monotone in the covariates and does not use a logistic link. Uniform overlap is guaranteed by construction.
\end{enumerate}

In all four designs the remaining coordinates \(x_4,\ldots,x_p\) are
irrelevant for treatment assignment. GenTE is trained on a half-sample
and evaluated by mean squared error against the oracle CATE on the
held-out fold, using the same architecture and the same number of
Monte Carlo replications as in Section \ref{sec:simulation}.

\begin{figure}[H]
\centering
\begin{subfigure}[t]{0.49\textwidth}
    \centering
    \includegraphics[width=\linewidth]{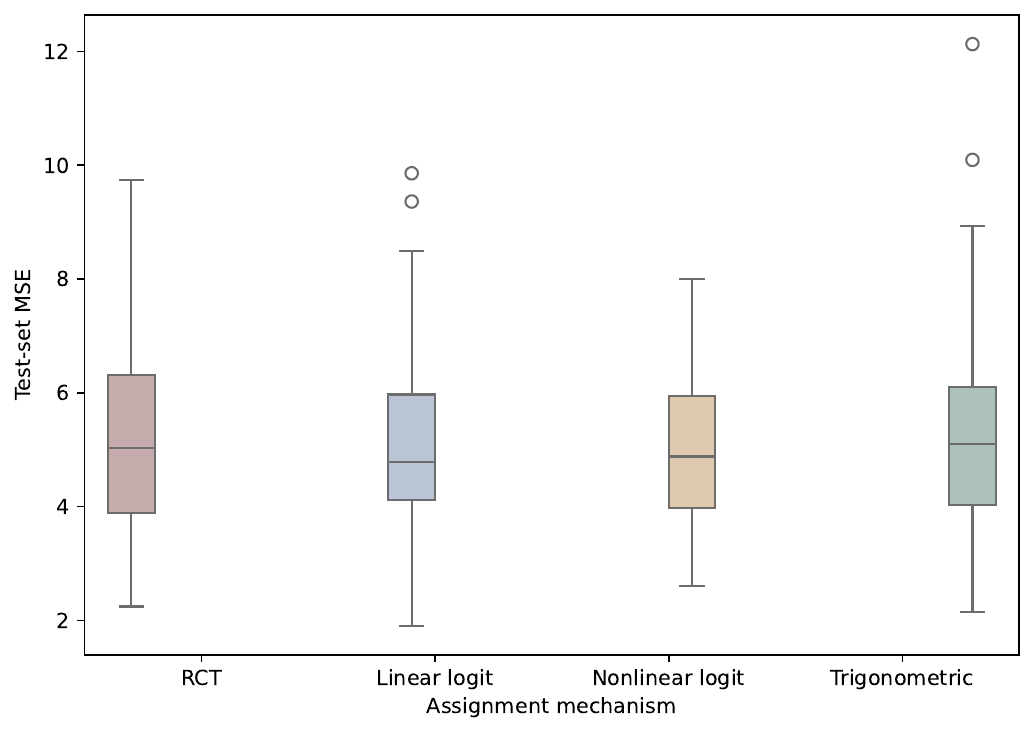}
    \caption{Out-of-sample mean squared error of the GenTE estimate of the conditional average treatment effect, evaluated against the oracle CATE under four treatment assignment mechanisms, with the CATE held fixed at the nonlinear interaction design. Each box summarizes 100 Monte Carlo replications. \(S = 500\) and \(p = 20\).}
    \label{fig:assignment_mse}
\end{subfigure}
\hfill
\begin{subfigure}[t]{0.49\textwidth}
    \centering
    \includegraphics[width=\linewidth]{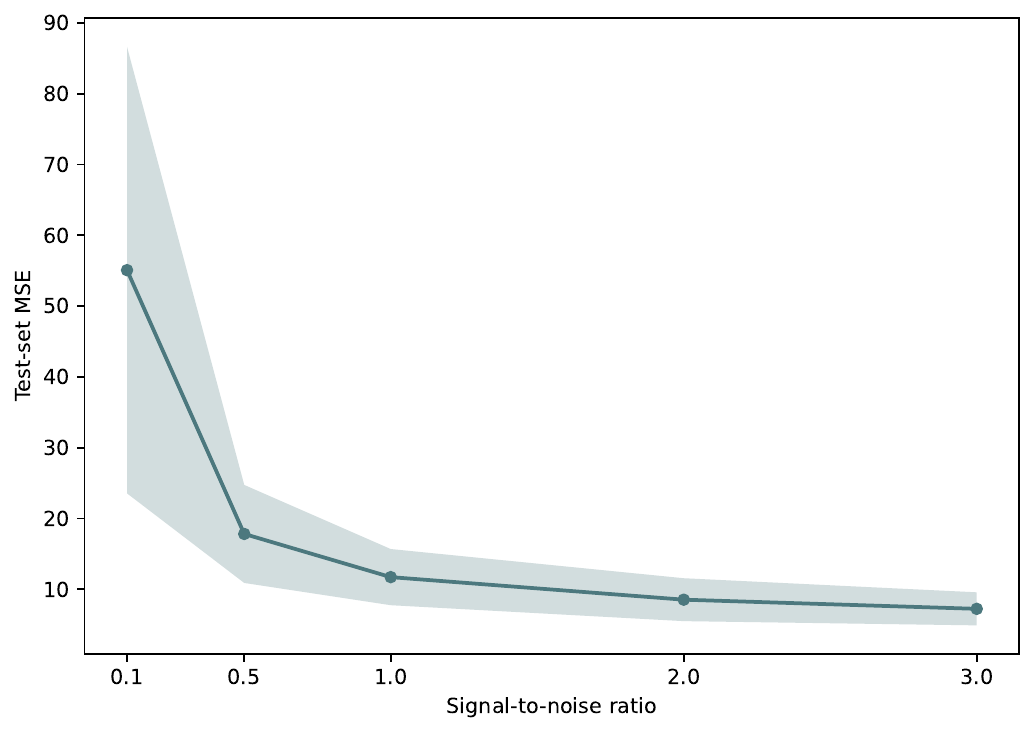}
    \caption{Out-of-sample mean squared error of the GenTE estimate of the conditional average treatment effect, evaluated against the oracle CATE, across values of the signal-to-noise ratio. The CATE is the nonlinear interaction design and assignment is the linear logit. The line is the mean MSE over 100 Monte Carlo replications. The ribbon is \(\pm\) one Monte Carlo standard deviation. \(S = 500\) and \(p = 20\).}
    \label{fig:snr_mse}
\end{subfigure}
\caption{Robustness of GenTE to the assignment mechanism (left) and to the signal-to-noise ratio (right).}
\label{fig:robustness}
\end{figure}

The boxplots in \autoref{fig:assignment_mse} show that GenTE's out-of-sample distribution of the 
mean squared error is essentially unchanged across the four assignment
mechanisms. With the CATE held fixed at the nonlinear interaction
design, $S=500$ and $p=20$, the Monte Carlo distributions under RCT,
linear logit, nonlinear logit, and trigonometric assignment have nearly
the same median, interquartile range, and upper tail. Once the outcome model is
fixed, replacing unconfounded randomization with a linear, nonlinear, or
non-monotone propensity does not materially affect the accuracy of the
estimator.

We next examine the robustness of the  performance across a range of signal-to-noise ratios.  Treatment assignment is the
linear logit:
\begin{equation}
  \pi(x)=\Lambda\bigl(0.3\,x_1-0.2\,x_2+0.1\,x_3\bigr),
  \qquad
  \Lambda(z)=(1+e^{-z})^{-1},
\end{equation}
and the CATE is held at the nonlinear interaction
\(\theta(x)=1+x_1^{2}x_2\). Covariates, baseline and residual remain
\begin{equation}
  X\sim\mathcal N(0,I_p),
  \qquad
  \mu(x)=x^{\top}\beta_{\mu},
  \qquad
  \beta_{\mu,j}\stackrel{\mathrm{iid}}{\sim}\mathcal N(0,1/4),
  \qquad
  \varepsilon\sim\mathcal N(0,\sigma^{2}),
\end{equation}
with \(p=20\). The outcome is
\(Y_s=\mu(X_s)+\theta(X_s)\,D_s+\varepsilon_s\). We define the signal-to-noise
ratio as the variance of the systematic part of \(Y_s\) relative to the
residual variance,
\begin{equation}
  \mathrm{SNR}
  =\frac{\mathrm{Var}\bigl(\mu(X_s)+\theta(X_s)D_s\bigr)}{\sigma^{2}}.
\end{equation}
The functions \(\mu\) and \(\theta\) are left unchanged, so the CATE
remains on the same scale as in the main simulations. Only
\(\sigma\) is recalibrated, using the sample variance of
\(\mu(X_s)+\theta(X_s)D_s\) in each draw, so that the target SNR is
attained. We consider
\(\mathrm{SNR}\in\{0.1,0.5,1,2,3\}\). As before, each simulated sample contains
\(S=500\) observations, split equally into training and evaluation sets. GenTE is trained on the first half and evaluated against the oracle CATE on the held-out half over 100 Monte Carlo replications.

\autoref{fig:snr_mse} shows that when the SNR lies in \(\{0.10,0.5\}\), the ribbon around the mean
test-set MSE is wide: residual variation dominates the systematic part
of \(Y_s\), and the estimator is unstable across replications. Once the
SNR exceeds \(0.5\), both the mean MSE and its Monte Carlo standard
deviation fall and then flatten, so that further increases in the
signal-to-noise ratio yield only modest gains in accuracy and precision.



\section{Error Distributions}\label{app_error_distributions}
This section compares two parameterizations of the same estimand. In the first, the treatment effect is a function of the covariates alone,
\(\theta(x)\), and is fitted by the mean squared error loss. In the second, it is indexed by the quantile level, \(\theta(x, q)\) fitted by the composite loss of \ref{app:loss}, and aggregated over \(q\) to recover \(\theta(x)\). To do so, we simulate four designs that hold the conditional average treatment effect and the
assignment mechanism fixed as in Subsection~\ref{sec_sims_mse} and vary only the conditional law of the
disturbances. For $X_s \sim \mathcal N(0, I_p)$ and
$\beta_{\mu,j}\stackrel{\mathrm{iid}}{\sim}\mathcal N(0, 0.5^2)$,
\[
Y_s = X_s^{\top}\beta_{\mu} + \theta(X_s)\,D_s + \varepsilon_s ,
\qquad
\theta(x) = 1 + x_1 + 0.5\,x_2 ,
\]
\[
D_s \mid X_s = x \;\sim\; \mathrm{Bernoulli}\bigl(\Lambda(0.3x_1 - 0.2x_2 + 0.1x_3)\bigr),
\]
with the third term omitted when $p = 2$. The disturbance is
\begin{align}\label{eq_noise_distrs}
\varepsilon_s \;\sim\;
\begin{cases}
\mathcal N(0, \sigma^{2}), & \text{(a) Gaussian},\\[4pt]
\sigma\,t_{3}/\sqrt{3}, & \text{(b) heavy-tailed},\\[4pt]
\sigma\,(W_s - m)/\sqrt{v}\ \ \text{if } D_s = 1, \quad
\mathcal N(0,\sigma^{2})\ \ \text{if } D_s = 0, & \text{(c) skewed},\\[4pt]
\mathcal N(0,\sigma^{2}) + \kappa\bigl(B_s - \varrho\bigr), & \text{(d) mixture},
\end{cases}
\end{align}
where $\sigma = 0.5$, $W_s \sim \mathrm{Lognormal}(0, s^{2})$ with $s = 0.8$,
$m = e^{s^{2}/2}$ and $v = (e^{s^{2}}-1)e^{s^{2}}$.  Moreover,
$B_s \sim \mathrm{Bernoulli}(\varrho)$ with $\varrho = 0.15$ and $\kappa = 3$.
Every disturbance has conditional mean zero, so $\theta(\cdot)$ is the
conditional average treatment effect in all four designs. Design (d) is bimodal.

The comparison in \autoref{tab:gente_quantile_direct} suggests a trade-off between the two parameterizations rather
than a uniform ranking. When the disturbances are Gaussian, squared loss is the
efficient criterion for a conditional mean, and the direct specification attains
the lower mean MSE in seven of the nine cells. The same holds under
$t_3$ disturbances rescaled to the same variance, so tail thickness alone does
not favor the check function. The ordering shifts as the conditional outcome
law diverges from a single symmetric mode: the direct specification leads in five
of nine cells when the treated arm is skewed, and in only one of nine under the
mixture design, where the density is bimodal. Within every block the quantile-indexed
specification is relatively stronger at $p = 2$ and at the smaller sample sizes,
and weaker at $p = 8$, which is consistent with the pattern in
\autoref{tab_mse}: the same network capacity must represent a surface in
$(x, q)$ rather than a function of $x$ alone, and that cost grows with the
covariate dimension. The two quadratures of the CATE integral give
indistinguishable results throughout, as in \autoref{tab:gente_sep_joint}.
Taken together, the quantile parameterization is preferable when the conditional
outcome distribution is multimodal or asymmetric, and costs little when it is
not, while the direct specification is the more accurate choice for the average
effect under Gaussian errors in higher covariate dimensions. This concerns the
conditional average effect alone: the direct specification returns \(\theta(x)\), and
$\theta(x,q)$ is unavailable from it at any $q$.

\begin{table}[H]
\centering
\caption{Mean squared error of the GenTE estimate of the conditional average
treatment effect under four error distributions, by sample size $S$ and
covariate dimension $p$. The effect is held fixed at
$\theta(x) = 1 + x_{1} + 0.5\,x_{2}$ and the assignment mechanism at the linear
logit of Subsection~\ref{sec_sims_mse}, so the blocks differ only in the
conditional law of the disturbances, which is described in
\eqref{eq_noise_distrs}. GenTE (quantile) is the specification of the
main text, with the CATE obtained by averaging $\hat\theta(x,q)$ over a grid of
$19$ levels and, in the second block, over $500$ uniform draws.
GenTE (direct) removes the quantile indexing: both heads are functions
of $x$ alone and the outcome is fitted by squared loss, with all other settings
held fixed, so the comparison isolates the quantile parameterization. Cells
report the mean MSE against the oracle CATE, with the Monte Carlo standard
deviation in parentheses, over $100$ replications. \bt{Dark blue} marks the
cells in which the direct specification attains the lower MSE when compared to quantile averaging.}
\label{tab:gente_quantile_direct}
\resizebox{0.8\textwidth}{!}{%
\begin{tabular}{lrrrrrrrrr}
\toprule
& \multicolumn{3}{c}{GenTE (quantile, grid)} & \multicolumn{3}{c}{GenTE (quantile, MC)} & \multicolumn{3}{c}{GenTE (direct, squared loss)} \\
\cmidrule(lr){2-4}\cmidrule(lr){5-7}\cmidrule(lr){8-10}
Sample size & $p{=}2$ & $p{=}4$ & $p{=}8$ & $p{=}2$ & $p{=}4$ & $p{=}8$ & $p{=}2$ & $p{=}4$ & $p{=}8$ \\
\midrule
\multicolumn{10}{l}{Gaussian errors} \\
\midrule
200 & 0.107 & 0.345 & 0.712 & 0.106 & 0.346 & 0.715 & 0.321 & \bt{0.337} & \bt{0.366} \\
  & (0.067) & (0.165) & (0.380) & (0.067) & (0.165) & (0.381) & (0.218) & (0.123) & (0.140) \\
500 & 0.064 & 0.222 & 0.409 & 0.063 & 0.223 & 0.410 & 0.100 & \bt{0.221} & \bt{0.245} \\
  & (0.046) & (0.138) & (0.232) & (0.046) & (0.138) & (0.233) & (0.050) & (0.071) & (0.062) \\
1000 & 0.053 & 0.170 & 0.302 & 0.053 & 0.171 & 0.303 & \bt{0.039} & \bt{0.113} & \bt{0.171} \\
  & (0.046) & (0.101) & (0.147) & (0.047) & (0.102) & (0.148) & (0.016) & (0.037) & (0.040) \\
\addlinespace
\midrule
\multicolumn{10}{l}{Heavy-tailed errors ($t_3$)} \\
\midrule
200 & 0.084 & 0.316 & 0.673 & 0.084 & 0.317 & 0.674 & 0.247 & \bt{0.288} & \bt{0.385} \\
  & (0.061) & (0.161) & (0.347) & (0.061) & (0.162) & (0.348) & (0.201) & (0.149) & (0.304) \\
500 & 0.052 & 0.213 & 0.366 & 0.052 & 0.214 & 0.367 & 0.092 & \bt{0.193} & \bt{0.222} \\
  & (0.051) & (0.162) & (0.202) & (0.051) & (0.163) & (0.203) & (0.078) & (0.109) & (0.121) \\
1000 & 0.048 & 0.158 & 0.284 & 0.048 & 0.159 & 0.285 & \bt{0.044} & \bt{0.109} & \bt{0.157} \\
  & (0.049) & (0.101) & (0.140) & (0.049) & (0.102) & (0.140) & (0.040) & (0.103) & (0.074) \\
\addlinespace
\midrule
\multicolumn{10}{l}{Skewed treated arm} \\
\midrule
200 & 0.103 & 0.334 & 0.747 & 0.103 & 0.335 & 0.749 & 0.297 & 0.353 & \bt{0.405} \\
  & (0.087) & (0.184) & (0.455) & (0.087) & (0.184) & (0.456) & (0.325) & (0.207) & (0.228) \\
500 & 0.061 & 0.214 & 0.403 & 0.060 & 0.215 & 0.405 & 0.186 & 0.221 & \bt{0.264} \\
  & (0.050) & (0.149) & (0.238) & (0.050) & (0.150) & (0.239) & (0.821) & (0.109) & (0.110) \\
1000 & 0.054 & 0.169 & 0.277 & 0.053 & 0.170 & 0.277 & \bt{0.043} & \bt{0.126} & \bt{0.180} \\
  & (0.051) & (0.103) & (0.137) & (0.051) & (0.104) & (0.138) & (0.052) & (0.065) & (0.088) \\
\addlinespace
\midrule
\multicolumn{10}{l}{Mixture errors} \\
\midrule
200 & 0.310 & 0.984 & 1.568 & 0.313 & 0.990 & 1.577 & 2.050 & 1.735 & \bt{1.222} \\
  & (0.284) & (0.773) & (0.635) & (0.290) & (0.772) & (0.638) & (1.931) & (1.024) & (0.598) \\
500 & 0.147 & 0.357 & 0.777 & 0.148 & 0.365 & 0.786 & 0.758 & 1.182 & 1.121 \\
  & (0.099) & (0.178) & (0.316) & (0.098) & (0.180) & (0.320) & (0.339) & (0.483) & (0.362) \\
1000 & 0.136 & 0.285 & 0.473 & 0.136 & 0.294 & 0.481 & 0.364 & 0.766 & 0.903 \\
  & (0.140) & (0.117) & (0.178) & (0.138) & (0.120) & (0.182) & (0.135) & (0.240) & (0.222) \\
\bottomrule
\end{tabular}%
}
\end{table}

\section{Stellar Characteristics}\label{stellar_characteristics}

\begin{figure}[H]
  \centering
  \begin{subfigure}[t]{0.49\textwidth}
    \centering
    \includegraphics[width=\linewidth]{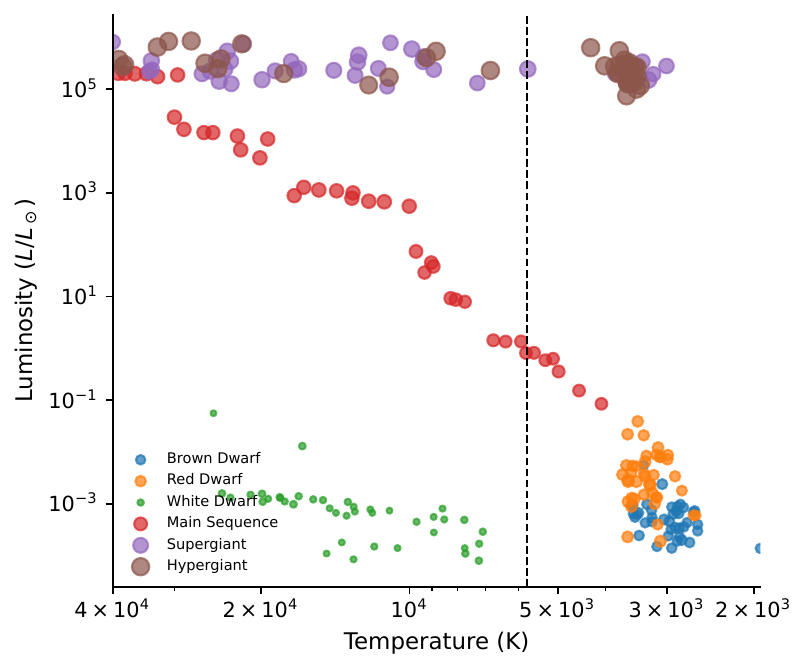}
    \caption{Hertzsprung--Russell diagram}
    \label{fig:star-hr}
  \end{subfigure}
  \hfill
  \begin{subfigure}[t]{0.49\textwidth}
    \centering
    \includegraphics[width=\linewidth]{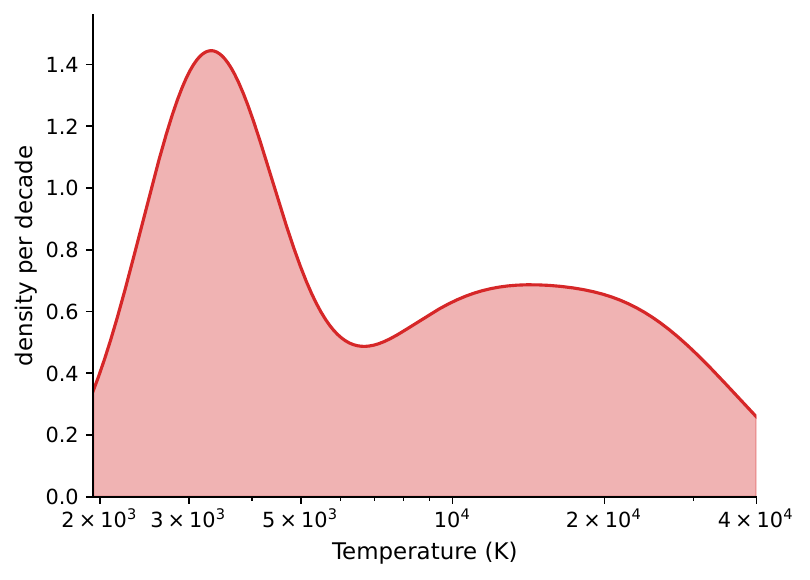}
    \caption{Temperature}
    \label{fig:star-density-T}
  \end{subfigure}
  \caption{Stellar characteristics. Panel~(a) is the Hertzsprung--Russell
diagram and panel~(b) the density of temperature, computed as a kernel density
of the base-10 logarithm so that the vertical scale is density per decade. The
dashed line in each panel is the sample median $t^{\star} = 5776$~K, which
defines the hot and cool temperature states.  Both axes
of panel~(a) are logarithmic and the temperature axis decreases from left to
right, following the usual convention for this diagram, whereas in panel~(b) it
increases. The figure uses the full catalog of $S = 240$ stars, before the
sample is split for estimation.} 
  \label{fig:star-temperature}
\end{figure}

\begin{figure}[H]
  \centering
  \begin{subfigure}[t]{0.49\textwidth}
    \centering
    \includegraphics[width=\linewidth]{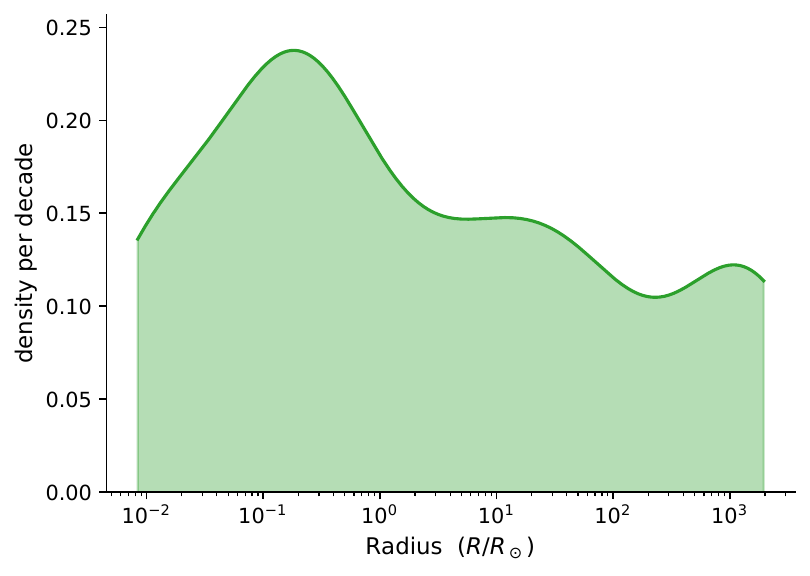}
    \caption{Radius}
    \label{fig:star-density-R}
  \end{subfigure}
  \hfill
  \begin{subfigure}[t]{0.49\textwidth}
    \centering
    \includegraphics[width=\linewidth]{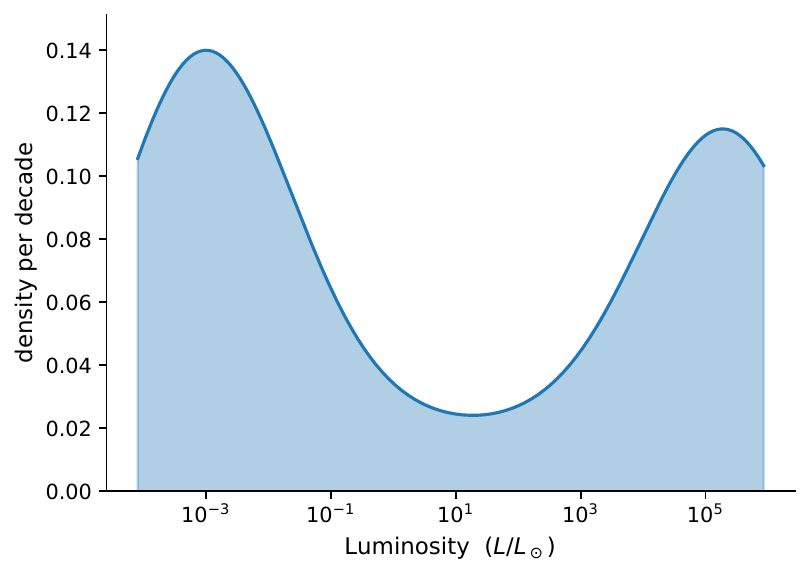}
    \caption{Luminosity}
    \label{fig:star-density-L}
  \end{subfigure}
  \caption{The densities of stellar radius (a) and luminosity (b). Both panels report a kernel density of the base-10 logarithm. The figure uses the full catalog of $S=240$ stars, before the
  sample is split for estimation.}
  \label{fig:star-size-luminosity}
\end{figure}

\section{Predicted Propensity Scores}\label{app_prop}

\begin{figure}[H]
\centering
\begin{subfigure}[t]{0.48\textwidth}
  \centering
  \includegraphics[width=\linewidth]{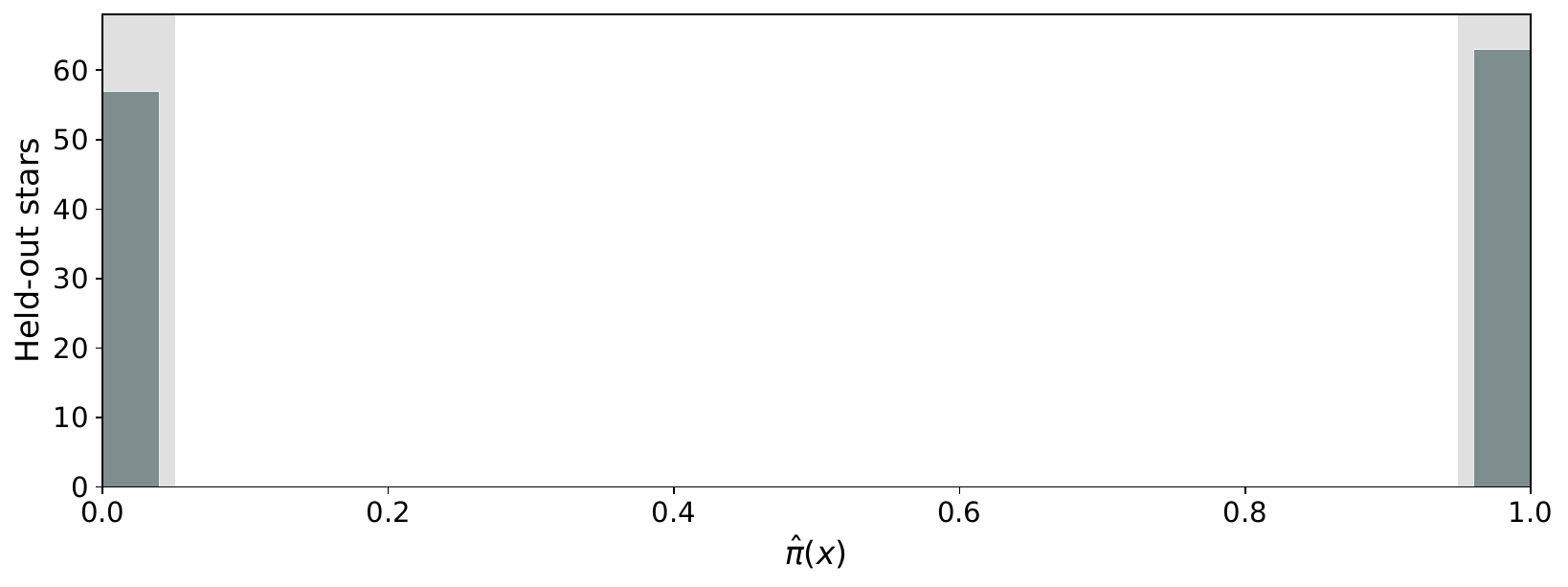}
  \caption{Limited empirical overlap}
\end{subfigure}\hfill
\begin{subfigure}[t]{0.48\textwidth}
  \centering
  \includegraphics[width=\linewidth]{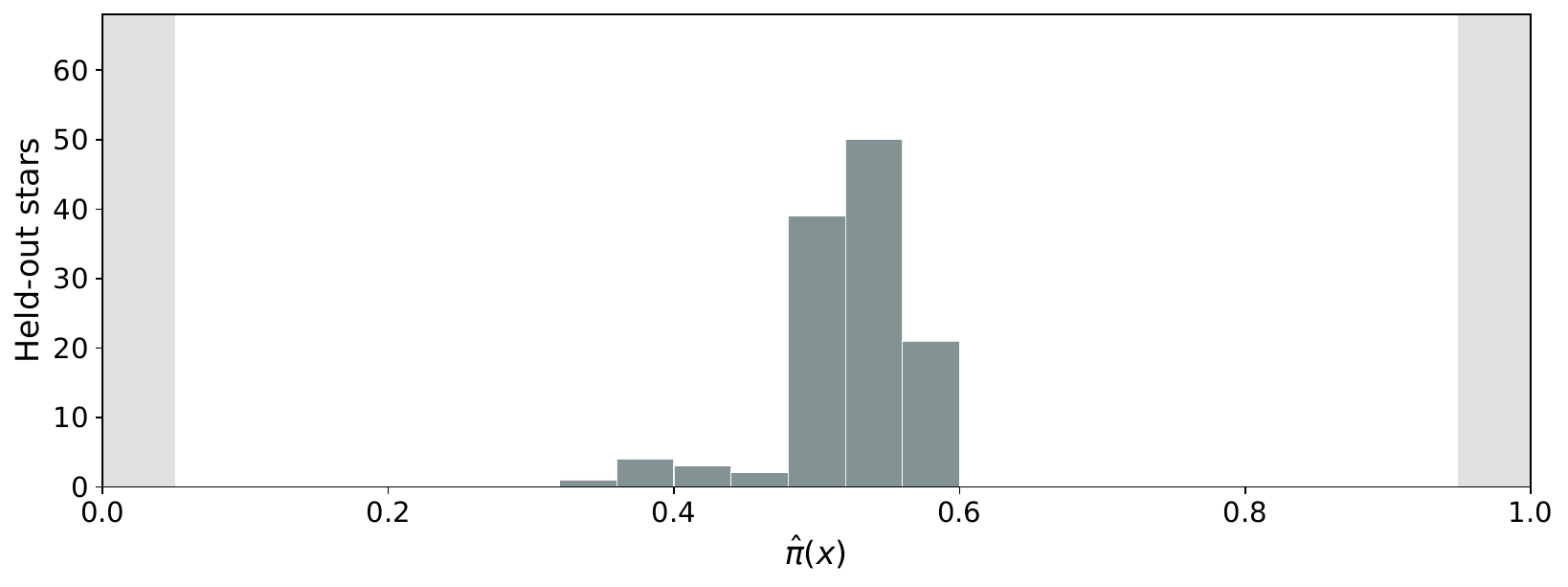}
  \caption{Improved empirical overlap}
\end{subfigure}
\caption{Estimated propensity scores on the held-out stars, from a
penalized logistic regression fitted on the training half.  The shaded
strips mark $\widehat{\pi}(x) < 0.05$ and $\widehat{\pi}(x) > 0.95$.
Under the full covariate set (stellar radius, color, type in panel (a)) all $120$ held-out stars fall in one of the
two strips.  Under radius and its
logarithm alone (panel (b)) every star lies in the interior.}
\label{fig:propensity_hist}
\end{figure}

\section{Predicted Stellar Luminosity}\label{app_pred_perf}

\begin{figure}[H]
\centering
\begin{subfigure}[t]{0.48\textwidth}
  \centering
  \includegraphics[width=\linewidth]{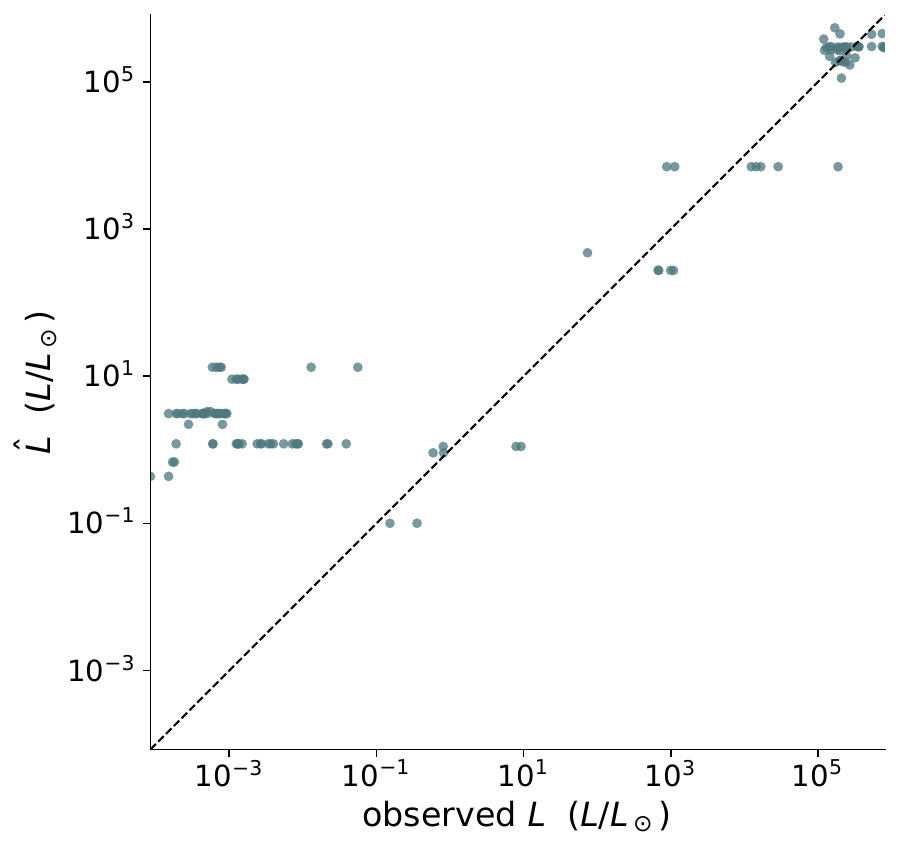}
  \caption{Limited empirical overlap}
  \label{fig:lfit_full}
\end{subfigure}\hfill
\begin{subfigure}[t]{0.48\textwidth}
  \centering
  \includegraphics[width=\linewidth]{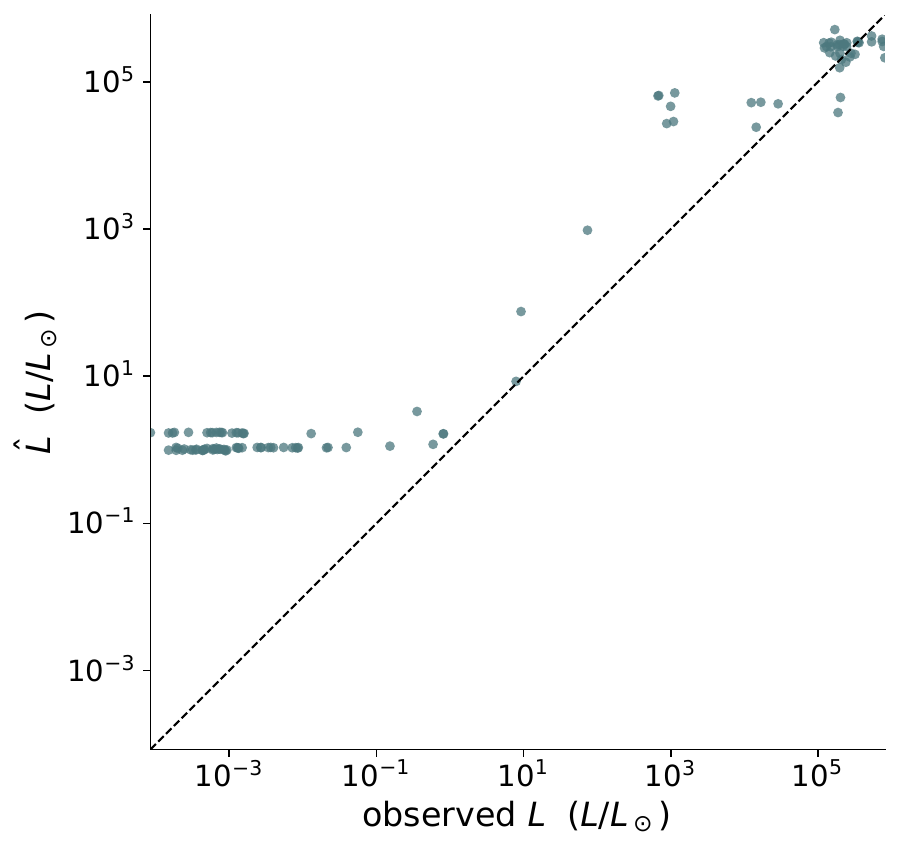}
  \caption{Improved empirical overlap}
  \label{fig:lfit_radius}
\end{subfigure}
\caption{Predicted against observed luminosity for the held-out stars,
in solar units.  The dashed line is the
$45^{\circ}$ line of perfect prediction.  Panel~(a) conditions on the
full covariate set, radius in level together with stellar
color and category, and panel~(b) on radius alone, in
level and logarithm.  Predictions track the observed values closely for
the luminous stars.  Below
roughly $10^{-2}\,L/L_{\odot}$ both specifications flatten: the fitted
values settle near $1\,L/L_{\odot}$ while the observed luminosities fall
by a further four decades.  These are the brown, red and white dwarfs,
for which the propensity score is degenerate, so the counterfactual arm
is unobserved and the fitted surface extrapolates rather than
interpolates.  The flattening is tighter in panel~(b), where radius is
the only covariate available to separate them.}
\label{fig:lfit_scatter}
\end{figure}

\end{appendix}

\end{document}